\documentclass[aps,prd,twocolumn,superscriptaddress,showpacs,amsmath,amssymb]{revtex4}
\usepackage{graphicx,epsf}

\usepackage[]{latexsym}
\newcommand{\be}{\begin{eqnarray}}
\newcommand{\ee}{\end{eqnarray}}

\def\comment#1{}

\begin{document}

%
%
\title{Glimpses on the micro black hole Planck phase}

%
%
\author{Fabio~Scardigli}
\email{fabio@yukawa.kyoto-u.ac.jp}
\affiliation{Yukawa Institute for Theoretical Physics, Kyoto
University, Kyoto 606-8502, Japan}
%
%
%
%
%
\begin{abstract}
Mass thresholds, lifetimes, entropy and heat capacity for micro black holes close to their
late Schwarzschild phase are computed
using two different generalized uncertainty principles, in the framework of models with
extra spatial dimensions. Emission of both photons and gravitons (in the bulk) are taken into account.
Results are discussed and compared.
\end{abstract}
\maketitle

\section{Introduction}\label{SEc1}

%
The appearance, some years ago, of scenar\^{\i} contemplating the existence of extra
space-like dimensions has had many different and deep impacts on the fundamental physics
of the last decade.

One of the most astonishing consequences of such framework seems to be the lowering
of the Planck scale to the TeV region, predicted by both the large extra dimensions
models \cite{ADD}, and by models with warped extra dimensional geometries \cite{RS}.
The explicit dependence of $M_{Planck}$ on the number of extra spatial dimensions varies
from a formulation to another, but all the formulations exhibit a value of $M_P$ which, at least
for some number of extra dimensions $n$, lies at the few TeV scale. This has had,
as an immediate consequence, the possibility to bring down to laboratory scale the "most wanted"
Holy Graal of modern theoretical physics, namely the Quantum Gravity regime. The unreachable
threshold of $10^{19}$ GeV (Planck energy in 4 dimensions) is reduced, in these frameworks,
to energies that are going to be vastly probed by the "next to start" CERN collider machine, LHC.
The center-of-mass energy created by this collider, $14$ TeV at full regime, implies that
we could have a (perhaps abundant) production of micro black holes.

Much work has been devoted to the understanding of the behaviour of these genuinely non
perturbative quantum gravity objects, from the seminal papers of \cite{SMRD, Emparan, G}
to the more recent analysis of the many possible signatures which could reveal these objects
in the not-so-far LHC runnings. See \cite{CGCRK} for a (very) incomplete list of papers
and further references.

The usual classification of the phases of the life of a micro black hole
(produced in a high energy collision) states that the decay of an excited
spinning black hole proceeds through several stages. There is a \emph{balding phase},
where the hole loses "hair" via the emission of classical gravitational and gauge radiation.
Gauge charges inherited from the initial colliding particles are being lost in this phase.
After, there is a brief \emph{rotating phase} (spindown phase) in which the black hole loses angular
momentum and energy, via super-radiance and semiclassical Hawking radiation, until it settles
down to a longer \emph{Schwarzschild phase}. Here, the black hole evaporates by loosing mass/energy
through the Hawking effect. According to the semiclassical approximation, the hole should
then shrink to zero, in a sort of final explosive state. At the very end point, the hole
should reach (according to semiclassical picture) an infinite temperature and an infinite
emission rate. Of course, it is widely believed that this semiclassical approximation
breaks down somewhere (well) before the Planck mass is reached.

While the first phases of the black hole life are intensely studied, and quite well understandable,
with the tools of QFT on curved space-time, the final part of the Schwarzschild phase, namely
the so called \emph{Planck phase}, is still by far the most puzzling sector of the problem.
We don't know much about this phase, mainly because we don't have a complete theory of
quantum gravity at our disposal \cite{GiddingsIC07}. However, some light on the
close proximity of the Planck phase (or, at least, on the last part of the Schwarzschild semiclassical phase)
seems to have been shred by several authors \cite{ACSantiago, CavagliaD, Hoss}.
They all use, as a good approximation of the final full quantum gravity theory, calculations
based on generalized uncertainty principle(s) (GUP), i.e. deformed commutators, in order to
compute the relevant quantities of the phenomenon, namely temperature, mass thresholds, black hole lifetimes.
As we shall see, a striking consequence of the GUP formalism is a direct prediction of the stopping
of the evaporation and the consequent existence of remnants. The present paper, which moves along the
research lines just above mentioned, is organized as follows: in Section II we introduce the two
generalized uncertainty principles we are dealing with; in Section III and Appendix 1 we prove the
translation and rotation invariance of such GUPs;
Sections IV and V are dedicated to the mass-temperature relation inferred by the GUPs and to the mass thresholds;
in Sections VI and VII we set up an emission rate equation, and use it to compute properties of the micro
black hole lifetime; Section VIII deals with the end point behaviour of the emission rate; in Section IX
we compute with the GUPs the entropy and the heat capacity of the micro black holes, and discuss the
important end point properties of the heat capacity. Section X is dedicated to the conclusions.

Throughout the paper, in $4+n$ dimensions, the Planck length is defined as $\ell_{4n}^{2+n}=G_{4n} \hbar /c^3$,
the Planck energy as $\mathcal{E}_{4n} \ell_{4n} = \hbar c /2$, the Planck mass as
$M_{4n}=\mathcal{E}_{4n}/c^2$. The value of the $4+n$ dimensional Newton constant $G_{4n}$, and its link
with the $4$ dimensional $G_N$, is in general model dependent
(see, e.g., \cite{G} for the ADD and RS model). A specific example will be discussed in Section V.

\section{GUPs}
%
Having in mind a high energy collision, we know that Heisenberg principle $\Delta p \Delta x \geq \hbar/2$
can be casted in the form $\Delta E \Delta x \geq \hbar c/2$ (since $\Delta E \simeq c\Delta p$).
It is worth noting that Heisenberg commutators in QFT  are Lorentz invariant statements,
which means that they are valid, in principle, at any energy scale. Actually, the main reason
since larger and larger energies are required to explore smaller and smaller details is that the size
of the smallest detail theoretically detectable with a beam of energy $E$ is $\delta x = \hbar c /(2 E)$.
An equivalent argument comes from considering the resolving power of a "microscope": the smallest
resolvable detail goes roughly as the wavelength of the employed photons, and therefore
$\delta x \simeq \lambda = \frac{c}{\nu} = \frac{hc}{\epsilon}$.

The research on viable generalizations of the Heisenberg uncertainty principle traces back to many
decades (see for early approaches \cite{GUPearly}, etc. See for a review \cite{GUPreview}
and for more recent approaches \cite{MM, FS}).
In the last 20 years, there have been seminal studies in string theory \cite{VenezETAL} suggesting
that in gedanken experiments about high energy scattering with high momentum transfer, the uncertainty
relation should be written as
\be
\delta x \,\gtrsim \,\frac{\hbar}{2p} \,+\, 2\beta \, \ell_{4n}^2 \,\frac{p}{\hbar}\, ,
\ee
where $\ell_{4n}$ is the $4+n$ dimensional Planck length. Since in our high energy scattering
$E \simeq c p$, the stringy GUP can be also written as
\be
\delta x \,\gtrsim \,\frac{\hbar c}{2 E} \,+\, 2\, \beta \,\ell_{4n}^2 \,\frac{E}{\hbar c}\, ,
\label{STGUP}
\ee
where $E$ is the energy of the colliding beams.\\
If however we take into account the possibility of a formation of micro black holes
in the scattering, with a gravitational radius of $R_S \sim (E)^{1/(n+1)}$, then we easily see that
in $4+n$ dimensions (and $n\geq1$) the stringy principle seems to forbid the very observation
of the micro hole itself. In fact, at high energy the error predicted by the stringy GUP goes
like $\delta x \sim E$, while the size of the hole goes like $R_S \sim (E)^{1/(n+1)}$.
For $E$ enough large and $n\geq1$, we always would have $E > (E)^{1/(n+1)}$, thereby
loosing the possibility of observing micro black holes just when they become massive (that is,
when they approach the classicality). Also to avoid this state of affairs, and on the ground of
gedanken experiments involving the formation of micro black holes, it has been proposed
\cite{FS, SC} a modification of the uncertainty principle that in $4+n$ dimensions
reads
\be \delta x \gtrsim \left\{ \begin{array}{ll}
\frac{\hbar c}{2 E} \quad {\rm for} \quad E < \mathcal{E}_{4n}\\ \\
\beta R_{4n}(E) \quad {\rm for} \quad E \geq \mathcal{E}_{4n} \, ,
\end{array}
\right.
\ee
where $R_{4n}$ is the $4+n$ dimensional Schwarzschild radius associated with the energy $E$
(see \cite{MyersPerry})
\be
R_{4n}=\left[\frac{16 \pi G_{4n} E}{(N-1) \Omega_{N-1} c^4}\right]^{\frac{1}{N-2}}
= \ell_{4n} \left(\omega_n \frac{E}{\mathcal{E}_{4n}}\right)^{\frac{1}{N-2}}
\ee
and $N=3+n$ is the number of space-like dimensions, $\omega_n=8\pi/((N-1)\Omega_{N-1})$,
$\Omega_{N-1} = 2 \pi^{N/2} / \Gamma(N/2) = {\rm area \,\,of \,\,the \,\,unit \,\,S^{N-1} \,\,sphere}$.
Combining linearly the above inequalities we get
\be
\delta x \gtrsim \frac{\hbar c}{2 E} \,+\, \beta R_{4n}(E)\, .
\ee
Thus, the GUP originating from micro black hole gedanken experiments (MBH GUP) can be written as
\be
\delta x \gtrsim \frac{\hbar c}{2 E} \,+\, \beta
\ell_{4n} \left(\omega_n \frac{E}{\mathcal{E}_{4n}}\right)^{\frac{1}{N-2}}\, .
\label{MBH}
\ee
Using again the relation $\mathcal{E}_{4n} \ell_{4n} = \hbar c /2$ the stringy inspired GUP
(ST GUP,  eq.(\ref{STGUP})) can be written in $4+n$ dimensions as
\be
\delta x \gtrsim \frac{\hbar c}{2 E} \,+\, \beta
\ell_{4n} \frac{E}{\mathcal{E}_{4n}}\, .
\label{ST}
\ee
where $\beta$ is the deformation parameter, generally believed of $O(1)$. Remarkably, in 4
dimensions ($N=3$, $n=0$) the two principles coincide. The deformation parameter $\beta$, supposed
independent from the dimensions $N$, can be therefore chosen as the same for both principles.\\
A comparison between the two principles yields
\be
\beta \ell_{4n} \frac{E}{\mathcal{E}_{4n}} >
\beta \ell_{4n} \left(\omega_n \frac{E}{\mathcal{E}_{4n}}\right)^{\frac{1}{N-2}}
 \Longleftrightarrow  E > (\omega_n)^{\frac{1}{n}} \mathcal{E}_{4n}
\ee
The last condition is easily met when $N=3,4,...,9$, in all the planned LHC collision
experiments, since $0.35 < (\omega_n)^{\frac{1}{n}} < 0.55$, and $\mathcal{E}_{4n}$ is believed to
be of order of few TeV (if extra dimensions are really there). As we mentioned, the above inequality
means that $\delta x_{ST} > \delta x_{MBH}$, so that at very high energies stringy GUP seems to produce
an error, $\delta x_{ST}$, larger then the size of the micro black hole itself. In other words the stringy GUP
does not seem to allow, in principle, for the observation of massive micro black hole (error larger than the
geometric size of the hole). This could be an indication supporting the MBH GUP (\ref{MBH})
in respect to ST GUP (\ref{ST}).

%
%
%
\section{Translation and rotation invariance of the GUPs}\label{SEc2c}
%
%
In this section we shall prove the GUPs previously introduced do respect the constraints
posed by requiring the conventional translation and rotation invariance of the commutation
relations. First, we show what these kinematic constraints imply about the structure, in $n$
dimensions, of the $[x,p]$ commutations relations. In this, we follow closely Ref. \cite{Kempf97_1, Kempf97_2}.
As a general ansatz for the $x, p$ commutation relation in $n$ dimensions we take
\be
[x_i, p_j] = i\,\hbar\,\, \Xi_{ij}(p)
\label{comm}
\ee
and we require that $\Xi_{ij}(p)$ differs significantly from $\delta_{ij}$ only for large momenta.
We assume also $[p_i,p_j]=0$ and we compute the remaining commutation relation through the Jacobi
identities, obtaining
\be
[x_i,x_j]=i\hbar \{x_a,\,\, \Xi_{ar}^{-1}\Xi_{s[i}\Xi_{j]r,s}\}
\label{comm2}
\ee
where $\{\}$ are the anti-commutators and $Q,_s:=\partial Q/\partial s$.
The commutation relations (\ref{comm}) are translation invariant (they are preserved
under the transformations $x_i \rightarrow x_i + d_i , \quad p_i \rightarrow p_i$).
However, the commutation relations (\ref{comm2}) are not invariant under translation, unless
we require $\Xi_{ij}(p)$ to be such that it yields $[x_i,x_j]=0$. Thus, in order to
implement translation invariance, $\Xi_{ij}$ must satisfy the necessary and sufficient condition
(read off from the (\ref{comm2}))
\be
\Xi_{ia}\partial_{p_i}\Xi_{bc}=\Xi_{ib}\partial_{p_i}\Xi_{ac}
\label{Tht}
\ee
where sum over $i$ is understood.
The rotation invariance can be implemented by requiring $\Xi_{ij}$ to have the form
\be
\Xi_{ij}(p)=f(p^2)\delta_{ij} + g(p^2)p_i p_j \, .
\label{Thr}
\ee
Together, conditions (\ref{Tht}) and (\ref{Thr}) imply that $f$ and $g$ must satisfy the
differential equation
\be
2 f' f + (2 f' p^2 -f)g = 0
\label{fg}
\ee
where $f'(p^2)=df/d(p^2)$. Under these conditions commutation relations do obey translation
and rotation invariance.

Considering, for sake of simplicity, the mono-dimensional case $i=j$, we write for the main
commutator
\be
[x, p] = i\hbar (f(p^2) + g(p^2)p^2)\, .
\label{fgcomm}
\ee
The usual Heisenberg commutator is recovered by choosing, for example, $f(p^2) = 1$. Then Eq.(\ref{fg}) implies
$g(p^2)=0$ and $[x,p]=i\hbar$. The stringy inspired commutator is obtained, to the first order in $\beta$,
by choosing $g(p^2)=\beta$ (see \cite{Kempf97_2}). Then, in fact, solving (\ref{fg}) we find
\be
f(p^2)=\frac{\beta p^2}{\sqrt{1+2\beta p^2}-1} \simeq 1 + \frac{\beta}{2}p^2 + O((\beta p^2)^2)
\ee
and, to the first order in $\beta$ (or, equivalently, for small $p$) we have
\be
[x,p]=i\hbar\left(1+\frac{3}{2}\beta p^2 + O(\beta^2)\right)\, .
\ee
The MBH GUP (\ref{MBH}) can be written in terms of momentum transferred as
\be
p \,\delta x \gtrsim \frac{\hbar}{2}\left(1 + \gamma p^{\frac{n+2}{n+1}}\right)
\ee
where
\be
\gamma = \beta (\omega_n)^{\frac{1}{n+1}}\left(\frac{2 \ell_{4n}}{\hbar}\right)^{\frac{n+2}{n+1}}
\ee
and this in terms of commutators becomes
\be
[x,p] = i \hbar \left(1 + \gamma p^{\frac{n+2}{n+1}}\right)\, .
\label{mbhcomm}
\ee
To show that MBH GUP is translation and rotation invariant we must show that the commutator (\ref{mbhcomm})
is of the same form of commutator (\ref{fgcomm}) (when $p \to 0$), with $f$ and $g$ satisfying (\ref{fg})
(in particular we would like to have $f(p^2) \to 1$ for $p \to 0$).
However, the previous strategy, namely to fix a priori a given form for $g(p^2)$ and then to compute $f(p^2)$
by solving (\ref{fg}) (as we did for HUP, $g(p^2)=0$, and for stringy GUP, $g(p^2)=\beta$),
in this case does not work. Even if one puts $p^2 g(p^2) = \gamma p^{(n+2)/(n+1)}$, Eq.(\ref{fg}) becomes however
rather complicated (it is a Abel equation of $2^{nd}$ kind), and hardly we can hope it gives $f(p^2) \to 1$ for
$p \to 0$. Moreover, an explicit solution could not be so useful, since we are mainly interested in
an asymptotic behaviour. Therefore we ask the following general properties to be satisfied by the functions
$f$ and $g$
\be \left\{ \begin{array}{ll}
[f(p^2) + g(p^2)p^2] \to [1 + \gamma p^{\frac{n+2}{n+1}}] \quad {\rm for} \quad p \to 0\\ \\
2 f' f + (2 f' p^2 -f)g = 0 \, ,
\end{array}
\right.
\label{condfg}
\ee
We shall look if there actually exist $f$ and $g$ such that the above two properties
can be simultaneously satisfied.
In this way the rotational and translational invariance of GUP (\ref{MBH}) will result proved.
In Appendix 1 such solutions are proved to exist, provided we allow $g$ to develop poles
(of course, the function $f$ and the whole function $f + g p^2$ remain perfectly finite).

%
%
%
%
\section{From the uncertainty principle to the mass-temperature relation}
%
%
%
Naturally, the modification of the uncertainty relation, i.e. of the basic commutators, has deep consequences
on the quantum mechanics, and on the quantum field theory, built upon it. The general implementation of such
commutation rules as regard Hilbert space representation, ultraviolet regularization, or modified dispersion
relations has been discussed in a vast literature (see for an incomplete list: \cite{vari}).
In the present section, we want to focus on the use of
(generalized) uncertainty relations to compute the basic feature of the Hawking effect, namely the formula
linking the temperature of the black hole to its mass $M$. The seminal result of Hawking and Unruh
\cite{Haw, Unr} is rigorously computed using QFT, based on Heisenberg uncertainty principle, on curved space-time.
However, it has been shown \cite{FS9506, ACSantiago, CDM03, CB05}
that the full calculation of QFT in
curved space-time (with standard commutators for the ordinary uncertainty principle, or with deformed
commutators for the GUP) can be safely replaced by a computation employing only the (generalized) uncertainty
relation and some basic physical considerations, in order to obtain the mass-temperature formula. \\
For example, in the case of the standard Hawking effect, we can consider a quantum of Hawking radiation
(a photon) just outside the event horizon of a Schwarzschild black hole of a given mass $M$.
Then, the uncertainty in the position of such quantum will be $\Delta x \simeq 2 R_S = 4GM/c^2$. The
correspondent uncertainty $\Delta E$ in the energy of the emitted quantum is identified with the thermal
energy of the quantum itself. For photons, the link between temperature and average thermal energy of photons
is $\Delta E \simeq 3 k_B T$. Therefore, the Heisenberg relation $\Delta x \Delta E \simeq \hbar c/2$ implies
\be
\frac{4GM}{c^2}\, 3 k_B T \simeq
\frac{\hbar c}{2} \quad \Rightarrow \quad T=\frac{\hbar c^3}{24 G k_B M}\,.
\ee
Note that the exact QFT coefficient is $8\pi$($\simeq 24$). A similar computation in Ref. \cite{ACSantiago},
based on the GUP in $4$ dimensions, has brought to a modification of the Hawking formula for
high temperatures, and to the remarkable prediction of black hole remnants.

Let us now consider, in $4+n$ dimensions, the two GUPs (\ref{ST}) and (\ref{MBH}) described in the Section II.
Suppose that, in a brane world scenario (ADD model or RS model)
a micro black hole with initial mass $M$ has been formed. Then, following the previous argument, let's consider
a quantum of Hawking radiation (photon or graviton) just outside the event horizon
(for simplicity, we suppose the black hole already in the Schwarzschild phase). The uncertainty in the
position of this quantum is $\delta x \simeq 2 R_{4n} = 2 \ell_{4n}(\omega_n m)^{\frac{1}{n+1}}$, where
$m=M/M_{4n}$. The correspondent uncertainty in energy of this quantum is identified with its thermal energy
$E$, and it is linked with the mass of the hole through the inequalities (\ref{ST}) or (\ref{MBH}),
now saturated. Reminding the relation $\mathcal{E}_{4n} \ell_{4n} = \hbar c /2$, from the GUPs formulae
we can write
\be
\label{ME}
2(\omega_n m)^{\frac{1}{n+1}} &=&
\frac{\mathcal{E}_{4n}}{E} \,+\, \beta \frac{E}{\mathcal{E}_{4n}} \quad \quad\quad\quad\quad\quad{\rm (ST)} \\
2(\omega_n m)^{\frac{1}{n+1}} &=& \frac{\mathcal{E}_{4n}}{E} \,+\, \beta
\left(\omega_n \frac{E}{\mathcal{E}_{4n}}\right)^{\frac{1}{n+1}}\quad  {\rm (MBH)}\nonumber \,.
\ee
Now, it is of fundamental importance to note that relations like the previous two can allow,
in principle, for arbitrary connections between energy $E$ and temperature $T$, without any affection of
the formulae for the minimum masses predicted by the relations themselves. In Appendix 2 we show that the
expression of the minimum mass predicted by Eqs.(\ref{ME}) is independent from the analytic structure
of the relation $E(T)$.\\
This freedom in the choice of $E(T)$ can be used to implement the correct semiclassical limit, in
$4+n$ dimensions, in relations like (\ref{ME}). Supposing still true a linear relation between $E$ and $T$,
$E=\alpha(n)T$, the correct form of $E(T)$ in $4+n$ dimensions can be inferred
by imposing a matching with the Hawking
semiclassical limit. In the limit $\beta \to 0$ we have, from (\ref{ME}),
\be
&&2(\omega_n m)^{\frac{1}{n+1}} = \frac{\mathcal{E}_{4n}}{\alpha(n)T} \quad \Rightarrow \nonumber \\
&&T = \frac{\mathcal{E}_{4n}}{2 \alpha(n)(\omega_n m)^{\frac{1}{n+1}}} =
\frac{\hbar c}{4 \alpha(n) R_{4n}}
\ee
and the matching with the correct Hawking semiclassical formula, in $4+n$ dimensions,
\be
T=\frac{(n+1)\hbar c}{4 \pi k_B R_{4n}}\, ,
\ee
can be obtained by setting
\be
\alpha(n)=\frac{\pi k_B}{n+1}\,.
\ee
Therefore $E(T)=\alpha(n)T=\pi k_B T/(n+1)$. Introducing the Planck temperature $T_{4n}$ defined by
$\mathcal{E}_{4n}=\frac{1}{2} k_B T_{4n}$, and using Planck units for the temperature itself
$\Theta = T/T_{4n}$, we can write from the (\ref{ME}) the mass-temperature relations as
\be
\label{MT}
&&2(\omega_n m)^{\frac{1}{n+1}} = \frac{(n+1)}{2 \pi \Theta} +
\beta \frac{2 \pi \Theta}{(n+1)} \quad \quad \quad{\rm (ST)} \\
\, \nonumber \\
&&2(\omega_n m)^{\frac{1}{n+1}} = \frac{n+1}{2 \pi \Theta} +
\beta \left(\frac{2 \pi \omega_n \Theta}{n+1}\right)^{\frac{1}{n+1}} \quad {\rm (MBH)}\nonumber
\ee
A possible source of ambiguity in relations like (\ref{MT}) is the relation between the uncertainty in the
position of the Hawking quantum, $\delta x$, and the geometric size of the hole, $R_{4n}$.
However, if we consider as valid a linear relation even with a free parameter $\mu$, like
$\delta x = 2 \mu R_{4n}$, we then see that
the Hawking limit can be recovered only by writing
\be
2 \mu (\omega_n m)^{\frac{1}{n+1}} = \mu \frac{n+1}{2 \pi \Theta}
\ee
and this, in the final mass-temperature relation, means
\be
2 \mu (\omega_n m)^{\frac{1}{n+1}} = \mu \frac{n+1}{2 \pi \Theta} + \beta \frac{2 \pi \Theta}{n+1}
\ee
which is always equivalent to a re-scaling of the unknown deformation parameter $\beta$
\be
2 (\omega_n m)^{\frac{1}{n+1}} = \frac{n+1}{2 \pi \Theta} + \left(\frac{\beta}{\mu}\right)
\frac{2 \pi \Theta}{n+1};
\quad \quad \beta'=\frac{\beta}{\mu}
\ee
Analogous considerations hold for the MBH GUP.

%
%
%
\section{Minimum masses, maximum temperatures}
%
%
%
The standard Hawking formula predicts a complete evaporation of a black hole, from an initial mass $M$
down to zero mass. As we have seen this is a direct consequence of the Heisenberg principle.
In the language of mass-temperature formula we have
\be
2 (\omega_n m)^{\frac{1}{n+1}} = \frac{n+1}{2 \pi \Theta}
\ee
and the temperature should become infinite at the very end of the process. Of course, as said in the
introduction, corrections are expected from quantum gravity processes in the final phases of the evaporation.
The GUPs seem to provide such corrections in a straightforward way. In fact, the mass-temperature formulae
(\ref{MT}) described in previous sections provide immediately a minimum mass for the evaporating black hole
and a maximum temperature. Precisely we have for the stringy GUP (reminding $N=n+3$, number of
space-like dimensions)
\begin{subequations}
\be
\Theta_{MAX}^{ST}=\frac{N-2}{2 \pi \sqrt{\beta}}
\ee
\be
m_{MIN}^{ST}=\frac{\beta^{\frac{1}{2}(N-2)}}{\omega_n}
\ee
\label{Thmax}
\end{subequations}
while for the MBH GUP we have
\begin{subequations}
\be
\Theta_{MAX}^{MBH}=\left( \frac{N-2}{2 \pi}\right)
\left[\frac{N-2}{\beta(\omega_n)^{\frac{1}{N-2}}}\right]^{\frac{N-2}{N-1}}
\ee
\be
m_{MIN}^{MBH}=\left[\frac{N-1}{2(N-2)^{\frac{N-2}{N-1}}}\right]^{N-2}
(\omega_n)^{-\frac{1}{N-1}}\beta^{\frac{(N-2)^2}{N-1}}
\ee
\label{ThmaxMBH}
\end{subequations}
Note that, as expected, $\Theta_{MAX} \to \infty$ and $m_{MIN} \to 0$ in the Hawking limit $\beta \to 0$.
Therefore the use of the GUP eliminates the problem of an infinite temperature (clearly un-physical) at the
end of the evaporation \cite{ACSantiago, CDM03, CavagliaD}.
In Section VIII we shall show that also the emission rate (erg/sec) is kept finite by the GUP
mass-temperature formulae, in contrast with an infinite output predicted by the Hawking formula (which is
based on Heisenberg uncertainty principle).
\begin{figure}[h]
\centerline{\epsfxsize=2.9truein\epsfysize=1.8truein\epsfbox{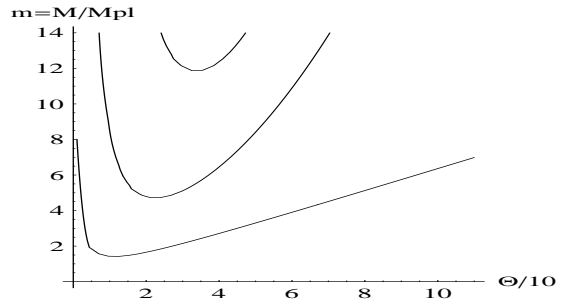}}
\caption[]{Mass - Temperature relation for the ST GUP, $\beta=2$, for $N=3,4,5$ from the bottom line to the top.}
\vspace{0.2cm} \hrule
\label{MassTempST}
\end{figure}

As it is clear from the mass-temperature diagrams (see Fig.\ref{MassTempST}),
black hole temperature is undefined for $m<m_{MIN}$.
Black holes with mass less than $m_{MIN}$ do not exist, since their horizon radius would fall below
the minimum allowed length. The mass-temperature relations derived from the GUPs predict hence the
existence of mass thresholds for the creation of micro black holes. An head-on collision of partons
with a center-of-mass energy below $m_{MIN}$ (in Planck units) will not result in production of black holes
(at least black holes as we define them today). It is interesting to compare the two diagrams
of $m_{MIN}^{ST}$ and $m_{MIN}^{MBH}$ as function of the number $N$ of space-like dimensions.
\begin{figure}[h]
\centerline{\epsfysize=1.8truein\epsfbox{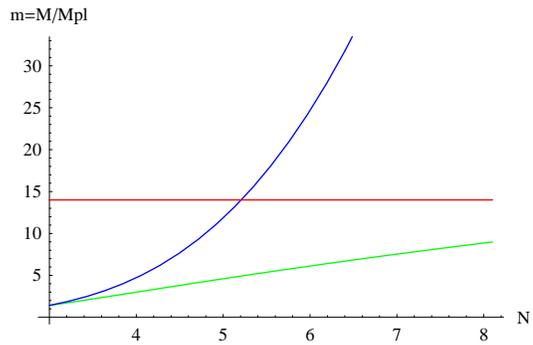}}
\caption[]{Minimal masses for ST (blue line) and MBH (green line) GUPs,
for $\beta=2$, and $M_{Pl}\simeq1$ TeV, for $N\geq4$. Red line is the LHC energy limit.}
\vspace{0.2cm} \hrule
\label{MasseMinime}
\end{figure}
In Fig.\ref{MasseMinime} these functions are plotted for $\beta =2$. On the y-axis we have the mass $m=M/M_{Pl}$ in
Planck units. If we assume that $M_{Pl} \simeq O(1)\,\,{\rm TeV}$ for $4\leq N \leq 9$,
as it is customary in many brane world models, then the diagram should be considered for $N\geq 4$ only.
(In fact, it is well known that for $N=3$ we have $M_{Pl} \simeq 10^{16}\,\,{\rm TeV}$, so surely LHC won't
create any black hole if the space-time dimensions are only $4$). If $M_{Pl} \simeq O(1)\,\,{\rm TeV}$, then
the horizontal red line represents the energy limit of LHC. We see that we have $m_{MIN}^{MBH} < m_{MIN}^{ST}$
for $N\geq 4$. In $N=4$ or $N=5$ both GUPs allow for the formation of black holes below the LHC threshold,
since $m_{MIN}<14$. On the contrary, already for $N=6$, and larger, only $m_{MIN}^{MBH}$ lies below the
LHC energy limit, while the stringy GUP predicts a $m_{MIN}^{ST}$ well above the energy reachable by LHC.
Therefore, productions of micro black holes in a scenario with $N\geq 6$ ($n\geq 3$ extra dimensions)
should be considered, if accompanied with remnants, an evidence in favor of the MBH GUP. \\
However, it is also true that for $0< \beta \lesssim 1.11$ both diagrams lie below the limit
of $14$TeV (for $4\leq N \leq 9$) so production of micro black holes is in principle allowed by both GUPs for
such values of $\beta$, although it is much more enhanced by MBH GUP.

As we know, the dependence of $M_{4n}$ from the dimensionality $N$ is different for different extra dimensions
scenarii. For example, in the ADD model \cite{ADD} the
extra dimensions have a finite size $L$. The link between $G_{4n}$ and the usual Newton constant $G_N$ is
$G_{4n} \simeq G_N L^{N-3}$ ($N=$ number of space-like dimensions). Therefore the relation among
$\mathcal{E}_{4n}$, the 4-dimensional Planck energy $\mathcal{E}_{Pl}$, and length $\ell_{Pl}$, is
\be
\mathcal{E}_{4n} = \left(\frac{\ell_{Pl}}{L}\right)^{\frac{N-3}{N-1}}\mathcal{E}_{Pl}\,.
\ee
Considering "large" extra dimensions of size $L = 1 \mu m = 10^{-4}$cm (and $\ell_{Pl}=1.6 \cdot 10^{-33}$ cm,
$\mathcal{E}_{Pl}=6.13 \cdot 10^{15}$ TeV), we have that in this model the Planck energy unit scales with
the number $N$ of dimensions like
\be
\mathcal{E}_{4n} = (1.6 \cdot 10^{-29})^{\frac{N-3}{N-1}}(6.13 \cdot 10^{15})\,\, {\rm TeV}
\ee
\begin{figure}[h]
\centerline{\epsfysize=1.8truein\epsfbox{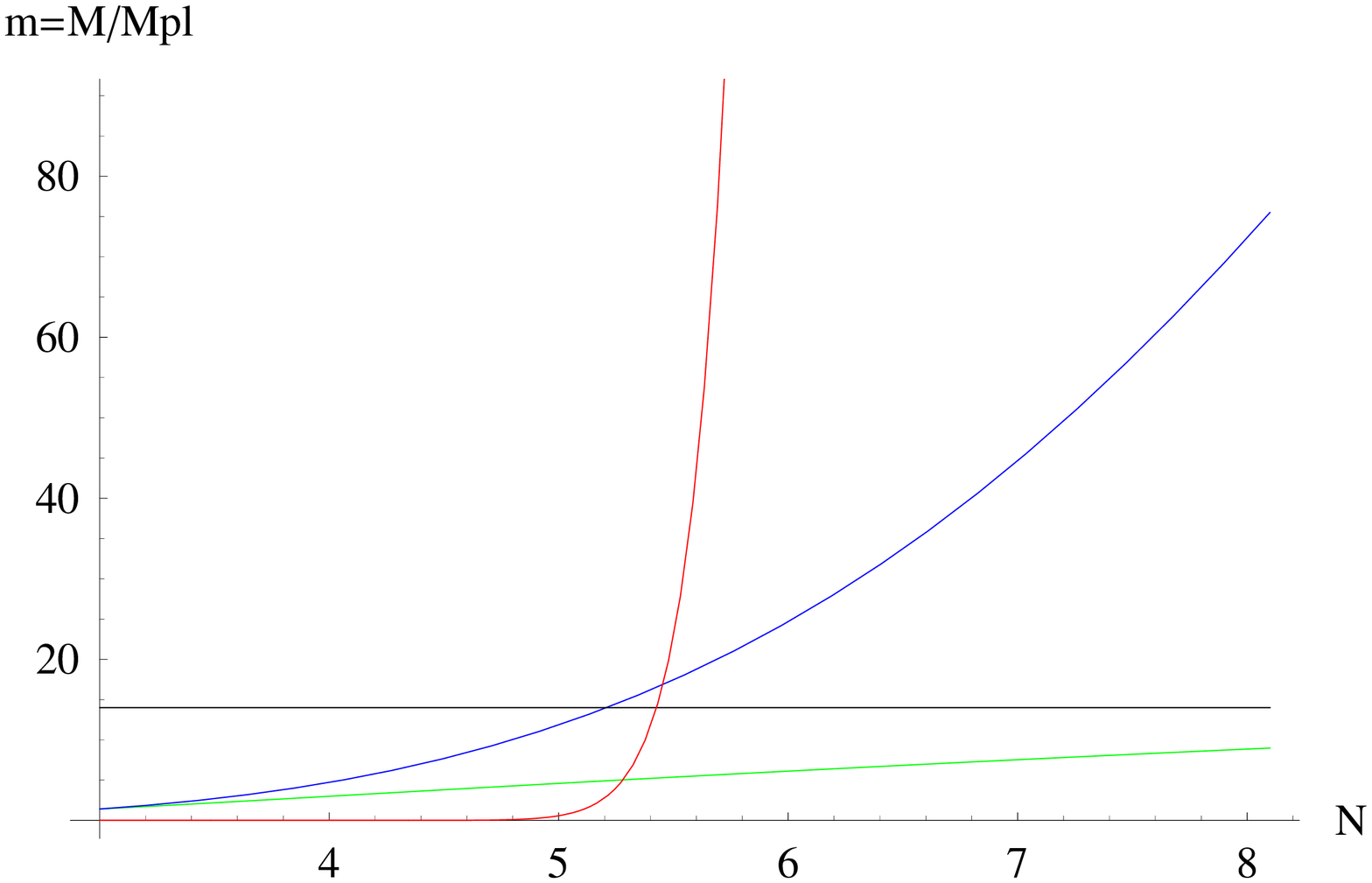}}
\caption[]{Minimal masses for ST and MBH GUPs, for $\beta=2$, with $M_{Pl}$ variable as in ADD model
(red line).}
\vspace{0.2cm} \hrule
\label{MasseMinimeADD}
\end{figure}
Fig.\ref{MasseMinimeADD} displays a diagram with the "limit-line" of $14$ TeV plotted in red in terms
of Planck mass units.
The energies reachable by LHC lie below the red curve. We see that in this particular ADD model there
is no hope to detect black holes for $N=4,\,5$, while for $N\geq6$ they should be detectable in any case,
and they are allowed by both GUPs, in that energy region.

However, the most striking prediction made by the GUPs mass-temperature relations (both versions) is,
without doubt, the existence of "remnants", i.e. of a final product of finite mass of the Hawking evaporation.
We shall show in Sections VIII and IX that Hawking evaporation should stop once the black hole mass reaches
$m_{MIN}$. The specific heat vanishes at the end point, so that the black hole cannot exchange heat
with surrounding environment, and the final object, a remnant, should be inert form the
thermodynamical point of view.
Remnants from primordial black holes have been proposed as a dark matter candidate \cite{ChenAd},
and they are also predicted by some models of quantum black holes \cite{barv}. \\
An obvious signature for revealing remnants at LHC collider would be the detection of a missing energy
of the order of the minimal mass of the black hole (plus missing energy due to invisible decay products).
Moreover, we see that both versions of GUP (stringy GUP  and MBH GUP) require the existence of remnants.
A detection of micro black holes, together with a non detection of remnants, would imply $\beta = 0$
(within experimental errors, of course). This would be a very strong assertion on the validity of the
usual (classical) Heisenberg principle down to any scale, and therefore a very deep insight on
the ultimate nature of the space time microstructure.
%
%
%
%
\section{Emission rate equation}
%
%
%
In this section we describe the evaporation process of the micro black hole. As it is well known,
in the minimal scenario of the braneworld models only gravity propagates also in the $4+n$ dimensional
bulk, whereas the Standard Model bosonic or fermionic fields are confined on a 4 dimensional brane.
In the present model we take into account two species of particles, photons and gravitons. This because
of several reasons: first, we wish to investigate typical examples of the two different geometries of emission;
second, some recent works \cite{CardCav, Cav03}
seem to show that the emission of gravitons in the bulk
may not be a negligible part of the total black hole emission, as previously believed
\cite{Emparan}; third, for sake of simplicity, we consider only photons as emitted
on the brane, in order to keep the model clear.
Nevertheless other kind of gauge or fermionic fields can be added in a straightforward way.

Before writing down the emission rate equation, we review some delicate issues about greybody
factors, emitted energy, and Stefan-Boltzamann constant, in $4$ and $4+n$ dimensions.

The number of photons (or gravitons) with frequency within $\omega$ and $\omega + d \omega$, in a volume $V$,
is given in $4$ dimensions by
\be
d n_{\gamma/g}=\frac{V \omega^2}{\pi^2 c^3}\,
\frac{\Gamma_{\gamma/g}(\omega)}{e^{\hbar\omega/k_BT}-1} \,d\omega\,.
\ee
In case of a perfect black body (perfect emitter) we have for the greybody factor
$\Gamma_{\gamma/g}(\omega) = 1$ for any $\omega$. The dependence of $\Gamma(\omega)$ from the frequency $\omega$
is in general very complicated. It has been studied in many papers (for 4 dimensional black holes see \cite{Page},
for emission of gravitons in $4+n$ dimensions see \cite{Cav03, CardCav}),
it is in some cases partially unknown, and in many cases can be computed only numerically. In the present model,
we neglect the dependence of the $\Gamma s$ from the frequency, and therefore we consider a value
$\Gamma_{\gamma/g}:=\langle\Gamma_{\gamma/g}(\omega)\rangle$ averaged on all the frequencies. Thus, for the number
of photons (or gravitons) in the interval ($\omega$, $\omega + d \omega$) in a volume $V$ we write
(in $4$ dimensions)
\be
d n_{\gamma/g}=\frac{V \omega^2}{\pi^2 c^3}\, \frac{\Gamma_{\gamma/g}}{e^{\hbar\omega/k_BT}-1} \,d\omega\,.
\ee
Obviously in a real black body (not the ideal one) will be $\Gamma_g \ll \Gamma_\gamma < 1$. \\
The total energy of photons (gravitons) contained in a volume $V$ in $4$ dimensions is then
\be
E^{\gamma/g}_{\rm TOT}(V) = \int_0^\infty \hbar \omega \,d n_{\gamma/g} =
\frac{V \Gamma_{\gamma/g}}{\pi^2 c^3 \hbar^3} (k_BT)^4 \Gamma(4)\zeta(4) \nonumber \\
\,
\ee
($\Gamma(x) = $ Euler Gamma function; $\zeta(n) = $ Riemann zeta function), and defining the Stefan-Boltzmann
constant in 4 dimensions as
\be
\sigma_3 = \frac{c}{3}\, \frac{\Gamma(4)\zeta(4)}{\pi^2 c^3 \hbar^3} k_B^4
\ee
this can be written
\be
E^{\gamma/g}_{\rm TOT}(V) = \Gamma_{\gamma/g}\, \frac{3 \sigma}{c}\, V T^4 \, .
\ee
In $4+n$ dimensions, only gravitons propagate in the bulk, and, taking into account their helicity,
we write for the number of gravitons in a volume $V$ and with frequency in the range
$\omega$, $\omega + d \omega$
\be
d n_g = \frac{(N+1)(N-2)}{2} \cdot \frac{V \Omega_{N-1} \omega^{N-1}}{(2 \pi c)^N} \times \nonumber \\
\times \frac{\Gamma_g}{e^{\hbar \omega / k_BT} - 1} \, d\omega
\ee
(where $N$ is the number of space-like dimensions).\\
The total energy of gravitons contained in a volume $V$ is then
\be
&&E^g_{\rm TOT}(V) = \int_0^\infty \hbar \omega \,d n_g = \frac{(N+1)(N-2)}{2} \times \nonumber \\
&&\times\frac{V \Gamma_g \Omega_{N-1} \hbar \Gamma(N+1)\zeta(N+1)}{(2 \pi c)^N}\,
\left(\frac{k_BT}{\hbar}\right)^{N+1}
\ee
which can be written as
\be
E^g_{\rm TOT}(V)=\Gamma_g \frac{N}{c} \,\sigma_N V T^{N+1}
\ee
where
\be
\sigma_N &=&\frac{c}{N}\frac{(N+1)(N-2)}{2}\frac{\Omega_{N-1}k_B^{N+1}}{(2\pi c)^N \hbar^N}\times\nonumber \\
&\times&\Gamma(N+1)\zeta(N+1)T^{N+1}
\ee
is the Stefan-Boltzmann constant in $N$ space-like dimensions ($4+n$ dimensions).

The total energy $dE$ radiated in a time $dt$, measured by the far observer,
from the black hole can be written (for photons and gravitons)
\be
d E = \Gamma_\gamma \frac{3\sigma_3}{c}\,\mathcal{V}_3\, T^4 \,+\,
\Gamma_g \frac{N\sigma_N}{c}\,\mathcal{V}_N\, T^{N+1}
\ee
where ${\cal V}_3$ is the effective volume occupied by photons
\be
{\cal V}_3 = \Omega_2\, R_{4n}^2\, c\, dt
\ee
and ${\cal V}_N$ is the effective volume occupied by gravitons
\be
{\cal V}_N = \Omega_{N-1}\, R_{4n}^{N-1}\, c\, dt \,.
\ee
Note that the Schwarzschild radius considered is always $R_{4n}$, i.e. the radius of the $4+n$ dimensional
black hole. This because we are dealing with micro black holes small enough than $R_{4n} < L$,
where $L$ is the typical size of the extra dimensions (in the ADD model), or the effective size of the
confinement of the zero mode of gravitational field (in the RS model).
Thus, finally, the differential equation of the
emission rate is \cite{SMRD, Emparan, Cav03}
\be
\frac{dE}{dt} &=& 3\, \Gamma_\gamma\, \sigma_3\, \Omega_2\, R_{4n}^2\, T^4 \nonumber \\
&+& N\, \Gamma_g\, \sigma_N\, \Omega_{N-1}\, R_{4n}^{N-1}\, T^{N+1}\, .
\label{ereq}
\ee
Using the explicit definitions of $\sigma_3$, $\sigma_N$, $R_{4n}$, and the Planck variables
$m=M/M_{4n}=E/\mathcal{E}_{4n}$, $\Theta=T/T_{4n}$, $\tau= t/t_{4n}$
(where $\mathcal{E}_{4n}=\frac{1}{2} k_B T_{4n}$ and $t_{4n}=\ell_{4n}/c$), we can rewrite
the emission rate equation as
\be
-\frac{dm}{d\tau} &=& \frac{2}{\pi^2}\,\Gamma_\gamma\,\Omega_2\,\Gamma(4)\,\zeta(4)\,
(\omega_n m)^{\frac{2}{N-2}}\,\Theta^4 \nonumber \\
&+& \frac{(N+1)(N-2)}{(2\pi)^N}\,\Gamma_g\,\Omega_{N-1}^2\,\times \nonumber \\
&\times&\Gamma(N+1)\,\zeta(N+1)\,(\omega_n m)^{\frac{N-1}{N-2}}\,\Theta^{N+1}
\label{ereqpl}
\ee
where the minus sign indicates the loss of mass/energy.
%
%
%
\section{Micro black hole lifetime}
%
%
%

In this section we study the emission rate equation in order to compute the micro
black hole lifetime. As a result, we shall display and comment various diagrams, in particular
for the limiting cases $\Gamma_\gamma\gg\Gamma_g$, $\Gamma_\gamma\ll\Gamma_g$.

The Eq.(\ref{ereqpl}) can be put in a simpler form by multiplying it by $\omega_n$ and defining
\be
y=(\omega_n m)^{\frac{1}{N-2}}\, .
\label{posy}
\ee
Then we have
\be
\frac{d(\omega_n m)}{d\tau} = (N-2) y^{N-3}\frac{dy}{d\tau}
\ee
and the Eq.(\ref{ereqpl}) becomes
\be
\label{ereqy}
&-&\frac{dy}{d\tau}=2\,\Gamma_\gamma\, \Omega_2\, \frac{\Gamma(4)\zeta(4)\omega_n}{\pi^2 (N-2)}\,\,y^{5-N}\,
\Theta^4 \\
&+& (N+1)\,\frac{\Gamma_g \omega_n \Omega_{N-1}^2 \Gamma(N+1)\zeta(N+1)}{(2\pi)^N}\,\, y^2\,
\Theta^{N+1} \, .\nonumber
\ee
Let us now examine separately the cases of the two different GUPs.\\
\\
\textbf{7a) Stringy GUP:} With the position (\ref{posy}), the stringy GUP can be written
\be
2 y = \frac{1}{\lambda \Theta} + \beta \lambda \Theta = \frac{1}{\bar{\Theta}} + \beta \bar{\Theta}
\label{stgupTh}
\ee
where $\lambda = 2 \pi /(N-2)$ and $\bar{\Theta}=\lambda\Theta$. Then (\ref{ereqy}) becomes
\be
-\frac{d(2y)}{d\tau} = \Gamma_\gamma\, G_n\, (2y)^{5-N}\,\bar{\Theta}^4\,
+\, \Gamma_g\, H_n\, (2y)^2\, \bar{\Theta}^{N+1}
\ee
with
\be
G_n = \frac{2^{N-3}\omega_n \Omega_2 \Gamma(4)\zeta(4)}{\pi^2 (N-2)\lambda^4}
\ee
and
\be
H_n = (N+1) \frac{\omega_n \Omega_{N-1}^2 \Gamma(N+1)\zeta(N+1)}{2(2\pi)^N \lambda^{N+1}} \, .
\ee
So we have a system of differential-algebraic equations
\be \left\{ \begin{array}{ll}
-\frac{d(2y)}{d\tau} = \Gamma_\gamma\, G_n\, (2y)^{5-N}\,\bar{\Theta}^4\,
+\, \Gamma_g\, H_n\, (2y)^2\, \bar{\Theta}^{N+1} \\ \\
2 y = \frac{1}{\bar{\Theta}} + \beta \bar{\Theta} \, .
\end{array}
\right.
\ee
This is easily reduced to a separable equation, in fact
\be
\mathcal{G}_N(\bar{\Theta}) d\bar{\Theta} = d \tau
\ee
where
\be
\mathcal{G}_N(\bar{\Theta}) =
\frac{(1 - \beta \bar{\Theta}^2)}
{[\Gamma_\gamma G_n (\beta \bar{\Theta}^2 + 1)^{3-N} + \Gamma_g H_n]
(\beta \bar{\Theta}^2 + 1)^2 \bar{\Theta}^{N+1}} \,. \nonumber
\ee
Suppose at time $\tau=0$ a micro black hole is created with an initial mass $m_0$ and
a (low) temperature $\bar{\Theta}_0$. At time $\tau$ the temperature has become $\bar{\Theta}(\tau)$,
with $\bar{\Theta}(0)=\bar{\Theta}_0$. Then
\be
\tau = \int_0^\tau d\tau =
\int_{\bar{\Theta}_0}^{\bar{\Theta}(\tau)}\mathcal{G}_N(\bar{\Theta}) d\bar{\Theta}\,.
\ee
By inverting this relation for $\bar{\Theta}$ one would get $\bar{\Theta}=f(\tau, \bar{\Theta}_0)$, that is
the temperature $\bar{\Theta}$ as a function of time $\tau$ and of the initial temperature $\bar{\Theta}_0$.
We are interested in the lifetime of the black hole. $\bar{\Theta}(\tau)$ grows with $\tau$ until a certain
value $\bar{\Theta}_{MAX}$, when the emission, as we shall see in Section IX, stops.
That value of $\tau$ will be the lifetime $\tau_{bh}$ of the black hole. So
\be
\tau_{bh}^{ST} = \int_{\bar{\Theta}_0}^{\bar{\Theta}_{MAX}}\mathcal{G}_N(\bar{\Theta}) d\bar{\Theta}\,.
\ee
It is easy to compute $\bar{\Theta}_{MAX}$ from Eq. (\ref{stgupTh}):
\be
\bar{\Theta}_{MAX} = \lambda\Theta_{MAX} = \frac{1}{\sqrt{\beta}}
\ee
which implies
\be
\Theta_{MAX} = \frac{N-2}{2 \pi \sqrt{\beta}}
\ee
in agreement with Eq. (\ref{Thmax}).\\
In order to obtain $\bar{\Theta}_0$ as a function of the initial mass $m_0$, as it is customary to do,
we have to invert the relation
\be
2 (\omega_n m_0)^{\frac{1}{N-2}} = \frac{1}{\bar{\Theta}} + \beta \bar{\Theta}
\ee
that is
\be
\beta \bar{\Theta}^2 - 2 (\omega_n m_0)^{\frac{1}{N-2}} \bar{\Theta} + 1 = 0 \, .
\ee
\begin{figure}[h]
\centerline{\epsfysize=1.8truein\epsfbox{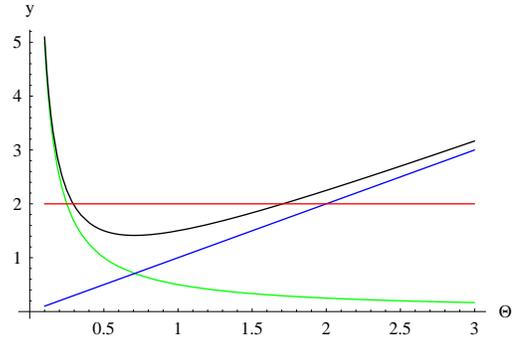}}
\caption[]{Diagram $y$ vs. $\bar{\Theta}$ for the stringy GUP, $\beta=2$.}
\vspace{0.2cm} \hrule
\label{GUPST}
\end{figure}
This can be done exactly ($2^{nd}$ degree equation). We choose the \emph{smallest positive} solution
(a simple glance to the diagram in Fig.\ref{GUPST} shows that such a solution always exists when $y > y_{MIN}$),
which means the solution of minimal temperature
\be
\bar{\Theta}_0 (m_0) = \frac{1}{\beta}\left[(\omega_n m_0)^{\frac{1}{N-2}} -
\left((\omega_n m_0)^{\frac{2}{N-2}} - \beta \right)^{\frac{1}{2}}\right].
\ee
Then we finally write for the lifetime of a micro black hole
\be
\tau_{bh}^{ST} = \int^{\bar{\Theta}_{MAX}}_{\bar{\Theta}_0(m_0)}\mathcal{G}_N(\bar{\Theta}) d\bar{\Theta}\,.
\label{taust}
\ee
In the following, we shall use numerical integration in order to produce some plots of the function
$\tau_{bh}^{ST}(m_0; N, \Gamma_\gamma, \Gamma_g, \beta)$ for several value of the
parameters $\Gamma_\gamma$, $\Gamma_g$.\\
\\
\textbf{7b) MBH GUP:} Here we proceed in an analogous way. Again with the position (\ref{posy}) the MBH GUP
can be written
\be
2 y = \frac{1}{\lambda \Theta} + \beta (\omega_n \lambda \Theta)^{\frac{1}{N-2}}
\label{mbhgupTh}
\ee
where $\lambda = 2 \pi /(N-2)$. Defining $\Lambda:=(\omega_n \lambda \Theta)^{\frac{1}{N-2}}$, which means
$\Theta = \Lambda^{N-2}/(\lambda\omega_n)$, then Eq.(\ref{mbhgupTh}) becomes
\be
2 y = \frac{\omega_n}{\Lambda^{N-2}} + \beta \Lambda \,.
\label{mbhgupLm}
\ee
Eq.(\ref{ereqy}) by virtue of $\Theta = \Lambda^{N-2}/(\lambda\omega_n)$ can be written
\be
-\frac{d(2y)}{d\tau} &=& \Gamma_\gamma\, \tilde{G}_n\, (2y)^{5-N}\,\Lambda^{4(N-2)} \nonumber \\
&+&\, \Gamma_g\, \tilde{H}_n\, (2y)^2\, \Lambda^{(N-2)(N+1)}
\ee
where now
\be
\tilde{G}_n = \frac{2^{N-3}\Omega_2 \Gamma(4)\zeta(4)}{\pi^2 (N-2)\lambda^4 \omega_n^3}
\ee
and
\be
\tilde{H}_n = (N+1) \frac{\Omega_{N-1}^2 \Gamma(N+1)\zeta(N+1)}{2(2\pi)^N \lambda^{N+1} \omega_n^N} \, .
\ee
Thus we have the system
\be \left\{ \begin{array}{ll}
-\frac{d(2y)}{d\tau} = \Gamma_\gamma\, \tilde{G}_n\, (2y)^{5-N}\,\Lambda^{4(N-2)} \\ \\
+\, \Gamma_g\, \tilde{H}_n\, (2y)^2\, \Lambda^{(N-2)(N+1)} \\ \\
2 y = \frac{\omega_n}{\Lambda^{N-2}} + \beta \Lambda \, .
\end{array}
\right.
\ee
Again, this can be reduced to a separable equation
\be
\mathcal{F}_N(\Lambda) d\Lambda = d\tau
\ee
where
\be
&&\mathcal{F}_N(\Lambda) = \nonumber \\
&&\frac{((N-2)\omega_n - \beta \Lambda^{N-1})}
{[\Gamma_\gamma \tilde{G}_n (\beta \Lambda^{N-1} + \omega_n)^{3-N} + \Gamma_g \tilde{H}_n]
(\beta \Lambda^{N-1} + \omega_n)^2 \Lambda^{(N-1)^2}} \,. \nonumber
\ee
We can now repeat the steps of the previous section (reminding that the change of variable
$\lambda \omega_n \Theta = \Lambda^{N-2}$ is a monotonic one). At time $\tau=0$ a micro black hole
with mass $m_0$ and temperature $\Theta_0$ ($\Lambda_0^{N-2}=\lambda\omega_n\Theta_0$) is created.
At time $\tau$, the temperature has grown to $\Theta(\tau)$, with
$\Lambda^{N-2}(\tau)=\lambda\omega_n\Theta(\tau)$ and $\Theta(0)=\Theta_0$, $\Lambda(0)=\Lambda_0$.
Then
\be
\tau = \int_{\Lambda_0}^{\Lambda(\tau)}\mathcal{F}_N(\Lambda) d\Lambda \,.
\ee
When $\Theta(\tau)$ reaches $\Theta_{MAX}$ (and $\Lambda(\tau)$ goes to $\Lambda_{MAX}$, respectively),
then the emission stops. $\Lambda_{MAX}$ can be computed as usual from (\ref{mbhgupLm}) and we get
\be
\Lambda_{MAX}=\left[\frac{(N-2)\omega_n}{\beta}\right]^{\frac{1}{N-1}} \,,
\ee
in agreement with (\ref{ThmaxMBH}).
$\Lambda_0$ as a function of the initial mass $m_0$ can be obtained as the smallest positive root of the
equation (numerical solution)
\be
\beta \Lambda_0^{N-1} - 2 (\omega_n m_0)^{\frac{1}{N-2}} \Lambda_0^{N-2} + \omega_n = 0 \,.
\ee
Again, a simple glance to the plot of Eq. (\ref{mbhgupLm}) assures us that such a solution always exists,
provided that $y > y_{MIN}$.

So, finally we can write for the lifetime of the micro black hole computed with the MBH GUP
\be
\tau_{bh}^{MBH} = \int^{\Lambda_{MAX}}_{\Lambda_0(m_0)}\mathcal{F}_N(\Lambda) d\Lambda \,.
\label{taumbh}
\ee
We can use numerical integration to produce some plots of the functions
$\tau_{bh}^{ST}(m_0; N,\Gamma_\gamma, \Gamma_g, \beta)$, $\tau_{bh}^{MBH}(m_0; N,\Gamma_\gamma, \Gamma_g, \beta)$
for several value of the parameters $\Gamma_\gamma$, $\Gamma_g$.

\begin{figure}[h]
\centerline{\epsfysize=1.8truein\epsfbox{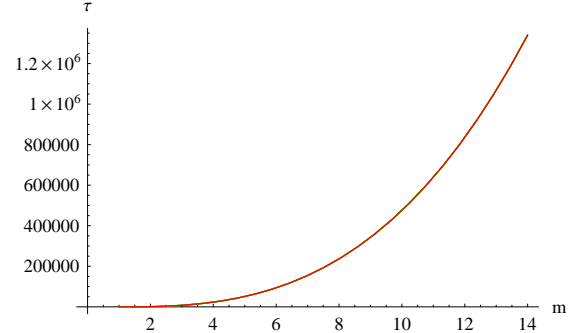}}
\caption[]{Diagrams of $\tau_{bh}$ for Stringy and MBH GUPs, for N=3 and $\beta=2$.
Mass and time in Planck units.}
\vspace{0.2cm} \hrule
\label{N=3mtauSTMBH}
\end{figure}

For example, for $N=3$ (the usual space-time), we see that the two GUPs, stringy and MBH, coincide, and so do
the two parts of the emission rate equation (photons and gravitons are both emitted on the brane, since
there isn't anything else outside). Also the mass threshold formulae of the two GUPs go to coincide.
Hence, we have one single diagram for the functions
$\tau_{bh}^{ST}(m_0; 3,\Gamma_\gamma, \Gamma_g, 2)$, $\tau_{bh}^{MBH}(m_0; 3,\Gamma_\gamma, \Gamma_g, 2)$,
as we see in Fig.\ref{N=3mtauSTMBH}.

Of course, Fig.\ref{N=3mtauSTMBH} is purely academical, since we know that for $N=3$ space-like dimensions,
the Planck
mass is about $10^{19}$ GeV, completely out of the energy range of LHC.

\begin{figure}[h]
\centerline{\epsfysize=1.8truein\epsfbox{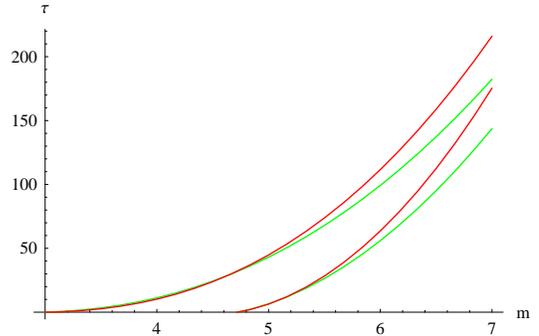}}
\caption[]{Diagrams of $\tau_{bh}$ vs. $m$. Green: emission of photons only. Red: emission of gravitons only.
Upper diagram obtained with MBH GUP, lower diagram with stringy GUP. $N=4$ and $\beta=2$.
Mass and time in Planck units (in $N=4$).}
\vspace{0.2cm} \hrule
\label{N=4mtauSTMBH}
\end{figure}

Instead, could be interesting Fig.\ref{N=4mtauSTMBH},
where $\tau_{bh}^{ST}$ and $\tau_{bh}^{MBH}$ are plotted for $N=4$. In green, we represent the emission
(on the brane) of photons only ($\Gamma_\gamma=1,\, \Gamma_g=0$); in red, the emission (bulk and brane)
of gravitons only ($\Gamma_\gamma=0,\, \Gamma_g=1$).
The upper diagram is obtained by the MBH GUP, the lower diagram represents the stringy GUP.
We note that the final mass predicted by the MBH GUP is lower than the final mass predicted
by the stringy GUP, as we already know from the formulae (\ref{Thmax}) and (\ref{ThmaxMBH})
and from Fig.\ref{MasseMinime}.

\begin{figure}[h]
\centerline{\epsfysize=1.8truein\epsfbox{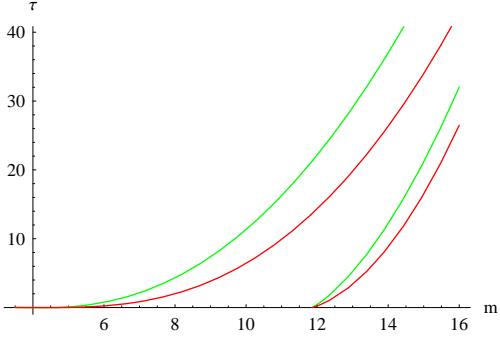}}
\caption[]{Diagrams of $\tau_{bh}$ vs. $m$. Green: emission of photons only. Red: emission of gravitons only.
Upper diagram obtained with MBH GUP, lower diagram with stringy GUP. $N=5$ and $\beta=2$.
Mass and time in Planck units (in $N=5$).}
\vspace{0.2cm} \hrule
\label{N=5mtauSTMBH}
\end{figure}

In Fig.\ref{N=5mtauSTMBH} we plot the diagrams describing the situation at $N=5$ (i.e. $n=2$ extra dimensions).
The gap
between the two mass' thresholds, predicted by the two different GUPs, has been further increased.
Note the switching between the photons and the gravitons emission lines (green and red) in respect to the
$N=4$ diagrams.

\begin{figure}[h]
\centerline{\epsfysize=1.8truein\epsfbox{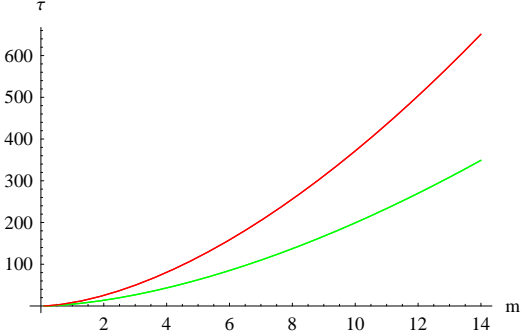}}
\caption[]{Diagrams of $\tau_{bh}$ vs. $m$ in the Hawking limit $\beta \to 0$.
Green: emission of photons only. Red: emission of gravitons only. $N=5$.
Mass and time in Planck units (in $N=5$).}
\vspace{0.2cm} \hrule
\label{HawkingLimitN=5}
\end{figure}

Finally, it is interesting to have a look to diagrams (Fig.\ref{HawkingLimitN=5}) representing
the Hawking limit $\beta \to 0$.
In such a limit the minimum masses go to zero, and obviously the two GUPs are going to coincide in between them,
and with the usual Heisenberg uncertainty principle.
%
%
%
%
%
\section{Emission rate equation at the end point}
%
%
%
%
In this section we show that the GUPs are able to keep finite the output rate of the black hole,
also at the end of the emission process. This, we shall see, is a completely different prediction
from the one made by the standard Hawking effect, which is in fact based on the usual Heisenberg
principle.

Before going any further, it is useful to expand the temperature $\Theta$ in series of the deformation
parameter $\beta$. In the case of the MBH GUP, since it is not possible to obtain an explicit form of
$\Theta(m)$, we shall use the implicit function theorem. The ST GUP case can be treated as well with the
implicit theorem, even if here $\Theta(m)$ can also be obtained, obviously, in an explicit form.\\
\\
\textbf{8a) Stringy GUP:} Rewriting the first of Eqs.(\ref{MT}) with the variables $y=(\omega_n m)^{1/(n+1)}$ and
$\bar{\Theta}=2\pi\Theta/(n+1)$, we have Eq.(\ref{stgupTh})
\be
2 y = \frac{1}{\bar{\Theta}} + \beta \bar{\Theta}
\label{ythbar}
\ee
The function $\bar{\Theta}=\bar{\Theta}(y,\beta)$ is implicitly defined by the equation
\be
f(y, \bar{\Theta}, \beta) = \beta\bar{\Theta}^2 - 2 y \bar{\Theta} +1 = 0
\ee
We are interested in $\bar{\Theta}$ as a function of $\beta$, therefore we consider $y$ to be constant.
Then
\be
df=0 \quad \Longrightarrow \quad
\frac{d\bar{\Theta}}{d\beta}=-\frac{\partial f / \partial \beta}{\partial f / \partial \bar{\Theta}}
\ee
and for the MacLaurin series of $\bar{\Theta}$ we get
\be
\bar{\Theta}(y,\beta) = \bar{\Theta}(y,0) + \frac{d\bar{\Theta}}{d\beta}(y,0)\cdot \beta + O(\beta^2)
\ee
with
\be
\bar{\Theta}(y,0)=\frac{1}{2y}; \quad \quad \quad \frac{d\bar{\Theta}}{d\beta}(y,0) = \frac{1}{8 y^3}
\ee
So finally
\be
\bar{\Theta}(y,\beta) = \frac{1}{2y} + \frac{\beta}{8 y^3} + O(\beta^2)
\label{svilth}
\ee
Writing this in term of the standard Hawking temperature $\bar{\Theta}_H = \frac{1}{2y}$ we have
\be
\bar{\Theta} = \bar{\Theta}_H (1 + \beta \bar{\Theta}_H^2 + O(\beta^2))
\ee
Using instead the usual variables $(m, \Theta)$ the expansion becomes
\be
\Theta = \frac{n+1}{4 \pi (\omega_n m)^{\frac{1}{n+1}}}
\left(1 + \frac{\beta}{4(\omega_n m)^{\frac{2}{n+1}}} + ...\right)\,.
\ee

\textbf{8b) MBH GUP:} The MBH GUP in $4+n$ dimensions, in the $(m,\Theta)$ variables, is given by the
second of Eqs.(\ref{MT}).
Here we \emph{should} use the implicit function method to obtain an expansion in $\beta$ of $\Theta(m,\beta)$.
Using the variables $y=(\omega_n m)^{1/(n+1)}$ and $\Lambda^{n+1} = 2\pi\omega_n\Theta/(n+1)$, the MBH GUP reads
\be
2y = \frac{\omega_n}{\Lambda^{n+1}} + \beta \Lambda
\label{yLa}
\ee
The function $\Lambda=\Lambda(y,\beta)$ is implicitly defined by the equation
\be
f(y,\Lambda,\beta) = \beta\Lambda^{n+2} - 2y\Lambda^{n+1} + \omega_n = 0
\ee
For MacLaurin at first order in $\beta$ we have
\be
\Lambda(y,\beta) = \left(\frac{\omega_n}{2y}\right)^{\frac{1}{n+1}} +
\frac{\beta}{2(n+1)y}\left(\frac{\omega_n}{2y}\right)^{\frac{2}{n+1}} + ...
\label{svilla}
\ee
Going back to the $(y,\Theta)$ variables, we have
\be
\Theta^{MBH} =
\frac{n+1}{4\pi y}\left[1+ \frac{\beta}{2y}\left(\frac{\omega_n}{2y}\right)^{\frac{1}{n+1}} + ...\right]
\label{expmbh}
\ee
while for the ST GUP in the same variables we had
\be
\Theta^{ST} =
\frac{n+1}{4\pi y}\left[1+ \frac{\beta}{4y^2} + ...\right]
\label{expst}
\ee
The results just obtained will be used later to construct an expansion in $\beta$ of the lifetime $\tau$.\\
\\
Let us now discuss some limit properties of the emission rate equation.\\
Starting from Eq.(\ref{ereqpl}) and introducing as usual the variable $y$ we arrive to Eq.(\ref{ereqy}),
that can be written as
\be
-\frac{dy}{d\tau} = A y^{2-n}\Theta^4 + B y^2 \Theta^{n+4}
\label{eryt}
\ee
with
\be
A=2\,\Gamma_\gamma\, \Omega_2\, \frac{\Gamma(4)\zeta(4)\omega_n}{\pi^2 (n+1)}
\ee
\be
B=(n+4)\,\frac{\Gamma_g \omega_n \Omega_{n+2}^2 \Gamma(n+4)\zeta(n+4)}{(2\pi)^{n+3}} \, .
\ee
In the limit $\beta \to 0$ the Hawking temperature can be recovered
\be
\Theta_H = \frac{n+1}{4 \pi y}
\ee
and the standard Hawking emission rate (in $4+n$ dimensions) reads
\be
-\left.\frac{dy}{d\tau}\right)_H &=& A y^{2-n}\Theta_H^4 + B y^2 \Theta_H^{n+4} \nonumber \\
&=& (\tilde{A}+\tilde{B})\frac{1}{y^{n+2}}
\label{sHaw}
\ee
where
\be
\tilde{A}:= A\left(\frac{n+1}{4\pi}\right)^4 ; \quad \quad \tilde{B}:= B\left(\frac{n+1}{4\pi}\right)^{n+4}\,.
\label{sHaw1}
\ee
We see now clearly that, in the case of the standard Hawking emission rate, we have an
infinite spike at the end of the evaporation process. In fact, when $m \to 0$, then $y \to 0$,
$\Theta_H \to \infty$, and $-\left.\frac{dy}{d\tau}\right)_H \to \infty$.

On the other hand, since the emission rate $\frac{dm}{d\tau}$ (or $\frac{dy}{d\tau}$) has to be a \emph{real}
number, we must require that the function $\Theta=\Theta(y,\beta)$ (implicitly defined by the GUPs relations)
is a real number.\\
A simple glance to the plots of the functions $(y,\Theta)$, for the ST and MBH GUPs, shows that $\Theta$
can be real if and only if $y > y_{MIN}$. In the case of ST GUP we have (from Eq.(\ref{stgupTh}))
$y_{MIN}^{ST}=\sqrt{\beta}$, while for the MBH GUP (from Eq.(\ref{mbhgupLm})) we get
\be
y_{MIN}^{MBH} = \frac{(n+2)\omega_n}{2}\left(\frac{\beta}{(n+1)\omega_n}\right)^{\frac{n+1}{n+2}}
\ee
Therefore, the request $\frac{dm}{d\tau} \in \mathbb{R}$ implies $y \geq y_{MIN}$, that is
$m \geq m_{MIN}$.
If now we compute the emission rate at the end point, when $m \to m_{MIN}$ (or $y \to y_{MIN}$) and
$\Theta \to \Theta_{MAX}$, we get, respectively,
\be
-\left.\frac{dy}{d\tau}\right)_{ST} = (\tilde{A}+\tilde{B}) \frac{1}{\beta^{1+\frac{n}{2}}}
\ee
\be
-\left.\frac{dy}{d\tau}\right)_{MBH} = (\tilde{E}+\tilde{F}) \frac{1}{\beta^{1+n}}
\ee
where $\tilde{A}, \tilde{B}, \tilde{E}, \tilde{F}$ are unimportant numerical factors.
Since $\beta > 0$, in both cases we see that the emission rate turns out to be \emph{finite} at the end point
of the evaporation. While the final stage of the standard Hawking process is cathastrophic (the black hole
reaches in a finite time a stage with zero rest mass, infinite emission rate, and infinite temperature),
the GUP (both versions) keeps all these quantities finite at the end of the evaporation. The modified results
appear to be more physically reasonable than the standard ones.\\
\\
\textbf{Expansion of lifetime $\tau$ in powers of $\beta$}: We conclude the section
presenting an expansion in series of $\beta$ of the emission rate equation, and the consequent
expansion of the lifetime $\tau$ of black holes. We use the variables $(y,\Theta)$.
From the Eqs.(\ref{expmbh}), (\ref{expst}) for $\Theta^{MBH}$ and $\Theta^{ST}$, and from
Eq.(\ref{eryt}) we can write, for the ST GUP
\be
\left.-\frac{dy}{d\tau}\right)_{ST} = A y^{2-n}\Theta^4 + B y^2 \Theta^{n+4} \nonumber \\
= (\tilde{A}+\tilde{B})\frac{1}{y^{n+2}}\left(1 + \beta \frac{\tilde{C}}{y^2} + ...\right)
\label{erexpst}
\ee
where $\tilde{A}$, $\tilde{B}$ are defined as in (\ref{sHaw1}) and
$\tilde{C} = (1+\frac{\tilde{B}}{\tilde{A}+\tilde{B}}\cdot \frac{n}{4})$.\\
For the MBH GUP we have, to the first order in $\beta$
\be
&&\left.-\frac{dy}{d\tau}\right)_{MBH} = \nonumber \\
&=& (\tilde{A}+\tilde{B})\frac{1}{y^{n+2}}
\left(1 + \frac{2\beta \tilde{C}}{y} \left(\frac{\omega_n}{2y}\right)^{\frac{1}{n+1}} + ...\right)
\label{erexpmbh}
\ee
with $\tilde{A}$, $\tilde{B}$, $\tilde{C}$ defined as before.\\
We are now enabled to develop the lifetime $\tau$ of the black hole in series of $\beta$.
Essentially, we integrate the expressions just found above for the emission rate
equation \footnote{An expansion in $\beta$ of the exact formulae (\ref{taust}), (\ref{taumbh}) for $\tau$
is in principle possible, but much more cumbersome than the method followed here.}.\\
\\
For the ST GUP, from (\ref{erexpst}) we have
\be
-d\tau = \frac{y^{n+2}}{\tilde{A}+\tilde{B}} \left(1 - \beta \frac{\tilde{C}}{y^2} + ...\right) dy
\ee
Taking the integral of the LHS between $0$ and $\tau$, and of the RHS between the initial
mass $m$ and the final mass
$m_{MIN}$ (i.e. between $y$ and $y_{MIN}=\sqrt{\beta}$), we have finally
\be
\tau &=& \frac{y^{n+3}}{(n+3)(\tilde{A}+\tilde{B})}
- \beta \frac{\tilde{C} y^{n+1}}{(n+1)(\tilde{A}+\tilde{B})}  +... \nonumber \\
&=& \frac{(\omega_n m)^{\frac{n+3}{n+1}}}{(n+3)(\tilde{A}+\tilde{B})}
- \beta \frac{\tilde{C}(\omega_n m)}{(n+1)(\tilde{A}+\tilde{B})} +...
\ee
The trend of $\tau$ with $m$ coincides with those found in Ref.\cite{CavagliaD}.\\
\\
For the MBH GUP, proceeding in an analogous way from Eq.(\ref{erexpmbh}), we find
\be
\tau &=& \frac{y^{n+3}}{(n+3)(\tilde{A}+\tilde{B})} \\
&-&2\beta \left(\frac{\tilde{C}(\omega_n/2)^{\frac{1}{n+1}}}{\tilde{A}+\tilde{B}}\right)
\frac{(n+1)}{[(n+1)^2+n]} y^{\frac{(n+1)^2+n}{(n+1)}} + ... \nonumber \\
&=& \frac{(\omega_n m)^{\frac{n+3}{n+1}}}{(n+3)(\tilde{A}+\tilde{B})} \nonumber \\
&-&2\beta \left(\frac{\tilde{C}(\omega_n/2)^{\frac{1}{n+1}}}{\tilde{A}+\tilde{B}}\right)
\frac{(n+1)}{[(n+1)^2+n]} (\omega_n m)^{1+\frac{n}{(n+1)^2}} + ... \nonumber
\ee
Once again we note that the zero order term ($\beta=0$) is the Hawking term (in $4+n$ dimensions)
and coincides with the that computed with ST GUP.
%
%
%
\section{Entropy and heat capacity}
%
%
%
%
In this section we compute the exact formulae for the thermodynamical entropy and for the heat capacity
of a (micro) black hole, using the two versions of the GUP previously introduced.
For sake of completeness, we shall give also an expansion in $\beta$ of the entropy $S$.
%
%
%
\subsection{Entropy}
%
%
From the first law of black hole thermodynamics \cite{BCH} we know that the differential of the thermodynamical
entropy of a Schwarzschild black hole reads
\be
dS = \frac{dE}{T_H}
\label{bht}
\ee
where $dE$ is the quantity of energy swallowed by a black hole with Hawking temperature $T_H$.
Rewriting Eq.(\ref{bht}) with the a-dimensional variables $(m, \Theta)$ we get
\be
d S = \frac{1}{2} k_B \frac{d m}{\Theta} \,.
\ee
Considering that $(m,\Theta)$ are linked by the GUP relation, we can in principle write $\Theta=\Theta(m)$,
and then obtain $S$ as a function of $m$ (mass of the micro hole), $S=S(m)$. This procedure is easily applicable
to the ST GUP, since $\Theta(m)$ can be found explicitly. However, this is not the case for the MBH GUP:
the function $\Theta(m)$ is determined only implicitly. Therefore, it is useful to express $m=m(\Theta)$
and then to arrive at $S$ as a function of $\Theta$, $S=S(\Theta)$. The integrals obtained in this way are also
more easily doable than those computed via the "$\Theta(m)$" method.\\
\\
Let us study the two cases separately, as usual.\\
\\
\textbf{9a) Stringy GUP:} Referring to formula (\ref{ythbar})  and using the variables $(y,\bar{\Theta})$
we have
\be
d S &=& \frac{1}{2} k_B \frac{d m}{\Theta} = \frac{\pi k_B}{\omega_n}\frac{y^n\,dy}{\bar{\Theta}} \nonumber \\
&=&\frac{\pi k_B}{2^{n+1}\omega_n}
\frac{(1 + \beta\bar{\Theta}^2)^n (\beta\bar{\Theta}^2 -1)}{\bar{\Theta}^{n+3}}\, d \bar{\Theta}
\ee
By integrating $dS$ we obtain $S=S(\bar{\Theta})$. We write the additive constant in $S$ so that
$S=0$ when $\Theta \to \Theta_{MAX}^{ST} = \frac{n+1}{2 \pi \sqrt{\beta}}$ (i.e. $\bar{\Theta} \to
\bar{\Theta}_{MAX}=\frac{1}{\sqrt{\beta}}$).
This is equivalent to what is usually done with the standard Hawking effect, where one fixes the additive constant
$=0$ for $m=0$ (the minimum mass attainable in the standard Hawking effect is $m=0$). Finally we can write
\be
S=\frac{\pi k_B}{2^{n+1}\omega_n} \int_{\bar{\Theta}_0(m_0)}^{\bar{\Theta}_{MAX}}
\frac{(1 + \beta\bar{\Theta}^2)^n (1 - \beta\bar{\Theta}^2)}{\bar{\Theta}^{n+3}}\, d \bar{\Theta}
\label{Sint}
\ee
where $\bar{\Theta}_0$ is the initial temperature. $\bar{\Theta}_0$ is fixed by the initial mass $m_0$
and can be expressed as a function of the initial mass $m_0$ by taking the smallest positive solution
of the equation
\be
\beta\bar{\Theta}_0^2 - 2 (\omega_n m_0)^{\frac{1}{n+1}} \bar{\Theta}_0 +1 = 0\,.
\ee
The integral (\ref{Sint}) can be easily solved analytically. For example in $4$ dimensions we have ($n=0$)
\be
dS = \frac{k_B}{4} \left(- \frac{1}{2\pi\Theta^3} + \frac{2\pi\beta}{\Theta}\right)\,d\Theta
\ee
and
\be
S(\Theta)=\frac{k_B}{16 \pi}\left(\frac{1}{\Theta^2} - \frac{1}{\Theta_{MAX}^2} +
8\pi^2\beta\log \frac{\Theta}{\Theta_{MAX}}\right)
\label{S4}
\ee
Note that $S \to 0^+$ for $\Theta \to \Theta_{MAX}^-$.\\
\\
\textbf{9b) MBH GUP:} Referring to formula (\ref{yLa})  and using the variables $(y,\Lambda)$
we have
\be
d S &=& \frac{1}{2} k_B \frac{d m}{\Theta} = \pi k_B \frac{y^n}{\Lambda^{n+1}}\,dy \\
&=&\frac{\pi k_B}{2^{n+1}}
\frac{(\beta \Lambda^{n+2} + \omega_n)^n (\beta\Lambda^{n+2} - (n+1)\omega_n)}{\Lambda^{(n+2)(n+1)+1}}\, d\Lambda
\nonumber
\ee
Integrating and choosing the additive constant so that $S \to 0$ when $\Theta \to \Theta_{MAX}$
(i.e. $\Lambda \to \Lambda_{MAX} = ((n+1)\omega_n/\beta)^{1/(n+2)}$) we have, finally,
\be
\label{Sintl}
&&S= \\
&&\frac{\pi k_B}{2^{n+1}}\int_{\Lambda_0(m_0)}^{\Lambda_{MAX}}
\frac{(\omega_n + \beta \Lambda^{n+2})^n ((n+1)\omega_n - \beta\Lambda^{n+2})}{\Lambda^{(n+2)(n+1)+1}}\, d\Lambda
\nonumber
\ee
where, as usual, $\Lambda_0$ is linked to the initial (low) temperature by
$\Lambda^{n+1}=2\pi\omega_n\Theta_0/(n+1)$ and can be expressed, if required, as function of the initial mass
$m_0$ by taking the smallest positive solution of the equation
\be
\beta\Lambda_0^{n+2} - 2(\omega_n m_0)^{\frac{1}{n+1}}\Lambda_0^{n+1} + \omega_n = 0
\ee
The integral (\ref{Sintl}) is analytically feasible, even if tedious. Of course, if we compute
Eq.(\ref{Sintl}) for $n=0$ (when the two GUPs coincide) we re-obtain Eq.(\ref{S4}).\\
\\
\textbf{Expansion of $S$ in $\beta$}. From a physical point of view we are interested also in the
series in $\beta$ of the entropy $S$.
Following a method already used for the series of lifetime $\tau$, we expand the differential form
$dS$ and then we integrate it.\\
In the case of ST GUP we have
\be
dS &=& \frac{\pi k_B}{\omega_n}\frac{y^n\,dy}{\bar{\Theta}} \nonumber \\
&=&\frac{2\pi k_B}{\omega_n}\left(y^{n+1} - \frac{\beta}{4} y^{n-1} + O(\beta^2) \right)\, dy
\ee
where variables $(y,\bar{\Theta})$ have been used, as well as formula (\ref{svilth}).
Adopting the usual normalization condition, that is $S=0$ for $m=m_{MIN}$
(i.e. for $y=y_{MIN}=\sqrt{\beta}$) and integrating we have
\be
S=\frac{2\pi k_B}{\omega_n}\int_{y_{MIN}}^y \left(y^{n+1} - \frac{\beta}{4} y^{n-1} + O(\beta^2) \right)dy
\ee
$\bullet$ for $n>0$
\be
S=\frac{2\pi k_B}{\omega_n}\left(\frac{y^{n+2}}{n+2} - \frac{\beta}{4n}y^n + O(\beta^{1+\frac{n}{2}})\right)
\ee
$\bullet$ for $n=0$
\be
\label{svils}
S&=&2\pi k_B\int_{y_{MIN}}^y \left(y - \frac{\beta}{4} y^{-1} + O(\beta^2)\,\alpha\, y^{-3} \right)dy \nonumber \\
&=&2\pi k_B \left[\frac{y^2}{2} - \frac{\beta}{4}\log y +
O(\beta^2)\frac{\alpha}{2y^2}\right]^y_{\sqrt{\beta}} \\
&=&2\pi k_B \left[\frac{y^2}{2} - \frac{\beta}{4}\left(\log y + 2 + K\right) +
\frac{1}{8}\beta\log\beta + O(\beta^2)\right] \nonumber
\ee
where $K$ is a \emph{numerical} factor which includes all the contributions from terms
of the form $O(\beta^{p+1})y^{-2p}$, $p=1,2,3,\dots$ computed in $y=\sqrt{\beta}$.
We see that the standard Bekenstein-Hawking entropy in $4+n$ dimensions is recovered in the limit $\beta \to 0$.
In particular, for $n=0$ we have $S \propto m^2$. Besides, we see that the leading correction to the standard BH
entropy induced by the GUPs is always negative: this means that the GUP-corrected entropy is smaller than the
semiclassical Bekenstein-Hawking entropy.\\
In the case MBH GUP, using the variables $(y,\Lambda)$ and referring to the formula (\ref{svilla}), we can write
\be
\Lambda^{n+1}(y,\beta) = \frac{\omega_n}{2y}
\left(1 + \frac{\beta}{2y}\left(\frac{\omega_n}{2y}\right)^{\frac{1}{n+1}} + \dots \right)
\ee
Therefore
\be
dS &=& \pi k_B \frac{y^n}{\Lambda^{n+1}}dy  \\
&=& \frac{2\pi k_B}{\omega_n}
\left(y^{n+1} -
\frac{\beta}{2}\left(\frac{\omega_n}{2}\right)^{\frac{1}{n+1}}y^{n-\frac{1}{n+1}} + \dots \right)dy
\nonumber
\ee
With the usual normalization condition ($S=0$ for $m=m_{MIN}$), we integrate from
\be
y_{MIN}=\frac{(n+2)\omega_n}{2}\left(\frac{\beta}{(n+1)\omega_n}\right)^{\frac{n+1}{n+2}} \nonumber
\ee
to a generic $y$, and obtain
\be
S&=&\frac{2\pi k_B}{\omega_n}\times \\
&\times& \int_{y_{MIN}}^y \left(y^{n+1} -
\frac{\beta}{2}\left(\frac{\omega_n}{2}\right)^{\frac{1}{n+1}}y^{n-\frac{1}{n+1}} + \dots \right)dy
\nonumber
\ee
$\bullet$ for $n>0$
\be
&S&=\frac{2\pi k_B}{\omega_n} \times \\
&\times&\left(\frac{y^{n+2}}{n+2} - \frac{\beta}{2}\left(\frac{\omega_n}{2}\right)^{\frac{1}{n+1}}
\left(\frac{n+1}{n(n+2)}\right) y^{\frac{n(n+2)}{n+1}} + \dots \right)     \nonumber
\ee
$\bullet$ for $n=0$\\
Here the results are exactly the same as for the ST GUP, formula (\ref{svils}),
and this is obvious, since for $n=0$ the two GUPs, ST and MBH, go to coincide.
Again, we note that the zero order, $\beta=0$, coincides with the usual Hawking entropy in $4+n$ dimensions
and with the value just previously obtained for the ST GUP.
%
%
%
\subsection{Heat Capacity}
%
%
%
In this section we compute the heat capacity of the (micro) black hole using the GUPs.
This quantity will give us important insights on the final stage of the evaporation process
(in particular, on the remnant state).\\
The heat capacity $C$ of a body is defined via the relation
\be
dE = C dT
\ee
meaning that the transfer to a body of an energy $dE$ produces a variation $dT$
in the temperature of the body itself.
Usually $dT > 0$, for usual bodies, and therefore $C>0$. As we know, this is not the case for the black holes.
Expressing the heat capacity via the variables $(m, \Theta)$ we get
\be
C=\frac{dE}{dT} = \frac{1}{2}k_B\frac{dm}{d\Theta}
\ee
So, for the ST GUP, using the variables $(y,\bar{\Theta})$ we can write
\be
C=\frac{\pi k_B}{2^{n+1}\omega_n}\frac{(1 + \beta\bar{\Theta}^2)^n(\beta\bar{\Theta}^2 - 1)}{\bar{\Theta}^{n+2}}
\ee
Since in general $0<\bar{\Theta}<\bar{\Theta}_{MAX}=\frac{1}{\sqrt{\beta}}$, from the above relation results $C<0$.
Note also that
\begin{itemize}
  \item If $\beta=0$, then $C<0$ for any $\Theta$. Black holes are bodies with negative specific heat.
  \item If $\beta=0$, then $C$ approaches $0$ ($C \to 0^-$) only when $\bar{\Theta} \to +\infty$.
  \item If $\beta>0$, then we have $C=0$ for $\bar{\Theta}=\bar{\Theta}_{MAX}=\frac{1}{\sqrt{\beta}}$, i.e.
  $\Theta=(n+1)/(2\pi\sqrt{\beta})$.
\end{itemize}
This means that if $\beta>0$ the specific heat vanishes at the end point of the evaporation process
in a finite time, so that the black hole
at the end of its evolution cannot exchange energy with the surrounding space. In other words,
the black hole stops to interact thermodynamically with the environment. The final stage of the Hawking
evaporation according to the GUP scenario contains a Planck-size remnant with a maximal temperature
$\Theta=\Theta_{MAX}$.\\
In particular, for $\beta=0$, we have, reintroducing the mass $m$,
\be
C=-\frac{2\pi k_B}{\omega_n}(\omega_n m)^{\frac{n+2}{n+1}}
\ee
\\
For the MBH GUP we use the variables $(y,\Lambda)$ and we have
\be
C &=& \frac{\pi k_B}{n+1}\frac{y^n}{\Lambda^n}\frac{dy}{d\Lambda} \\
&=&\frac{\pi k_B}{2^{n+1}(n+1)}\frac{1}{\Lambda^n}\left(\frac{\omega_n}{\Lambda^{n+1}} + \beta\Lambda\right)^n
\left(\beta - \frac{(n+1)\omega_n}{\Lambda^{n+2}}\right) \nonumber
\ee
If $\beta>0$ then $C=0$ when $\Lambda=\Lambda_{MAX}=\left[\frac{(n+1)\omega_n}{\beta}\right]^{\frac{1}{n+2}}$,
which means, again, that the heat capacity vanishes at the end point and the black hole is then thermodynamically
inert. When $\beta=0$ it is easy to verify that the above $C$ coincides with the one computed in the previous
paragraph, namely the case of standard Hawking effect in $4+n$ dimensions.
%
%
%
%
%
\section{Conclusions and outlooks}
%
%
%
In this paper we have examined the consequences that a deformed uncertainty principle has upon
relevant properties (mass threshold, lifetime, entropy, heat capacity) of micro black holes,
which could be produced at LHC in the (next) future (or could have been produced in the early universe,
or in cosmic rays showers), in the framework of models with extra spatial dimensions.

We have considered two possible deformations of the usual Heisenberg principle: one coming
from scattering gedanken experiments in string theory (shortly named ST GUP), the other coming
from gedanken experiments involving the formations of micro black holes (named MBH GUP).
A comparison of the basic predictions of the two principles at high energies seems to favor the
MBH GUP as more realistic. However, throughout the paper we have computed, and compared,
the consequences of both principles, using the same formalism, so that ultimately the
forthcoming \emph{actual} experiments could be the last judge of the predictions made by the two principles.
In respect to the previous literature on the subject we have considered also a non negligible
emission of gravitons in the $4+n$ dimensional bulk. Besides, the MBH GUP was not treated in precedent
literature.

The main conclusions of the paper can be summarized as follows:
\begin{itemize}
  \item Both principles predict remnants of finite rest mass as end product of Hawking evaporation of black holes.
  \item For deformation parameter $\beta \simeq 1.5$ or greater the mass thresholds predicted by MBH GUP
  are remarkably lower than those of ST GUP, meaning that the production of micro black hole is largely
  enhanced by the MBH GUP. In particular, micro black holes should be detectable in any number of extra
  dimensions, at the designed energy for LHC.
  \item The micro black hole lifetimes predicted by MBH GUP are in general always longer that those
  predicted by the ST GUP (and the difference is particularly noticeable for $N=5,6,...$ spatial dimensions).
  However the lifetimes predicted by both GUPs are, roughly, one order of magnitude shorter than those
  predicted by the standard Hawking evaporation (based on the standard Heisenberg principle)
  (see Fig.\ref{N=5mtauSTMBH} and Fig.\ref{HawkingLimitN=5}).
  \item The GUP-corrected entropy, for both GUPs, is lower than the standard Hawking entropy.
  \item The heat capacity predicted by both GUPs goes to zero at a finite (large) temperature,
  meaning that the black hole ceases then to interact thermodynamically with the environment. This is,
  in the present framework, a strong indication for the existence of an evaporation final product of
  finite mass, the remnant.

\end{itemize}
%
%
%
%
%
%
%
%
%
%
%
%
\section*{Acknowledgements}

F.S. thanks JSPS for support under the fellowship P06782.
%
%
%
%
\section*{Appendix 1}
%
In the differential equation (\ref{fg}) everything is function of $p^2$ and $f'(p^2)=df/d(p^2)$.
Then, setting $y:=p^2$ ($y>0$, $p=y^{1/2}$) we can write the two conditions (\ref{condfg})(necessary
and sufficient for translation and rotation invariance) as
\be \left\{ \begin{array}{ll}
[f(y) + g(y)y] \to [1 + \gamma y^{\frac{n+2}{2(n+1)}}] \quad {\rm for} \quad y \to 0\\ \\
2 f'(y) f(y) + 2y f'(y)g(y) - f(y)g(y) = 0 \, ,
\end{array}
\right.
\ee
To avoid fractionary powers, let's set $y^{\frac{1}{2(n+1)}}=:\lambda$, $y=\lambda^{2(n+1)}$. Then
\be
f(y) + g(y)y &=& f(\lambda^{2(n+1)}) + g(\lambda^{2(n+1)})\lambda^{2(n+1)} \nonumber \\
&=:& F(\lambda) + G(\lambda)\lambda^{2(n+1)}\, ,
\ee
Besides,
\be
f'(y)=\frac{df}{dy}=\frac{d\lambda}{d y}\frac{d F}{d \lambda} =
\frac{1}{2(n+1)}\lambda^{-(2n+1)}F'(\lambda) \, .
\ee
Thus the system becomes
\be \left\{ \begin{array}{ll}
[F(\lambda) + G(\lambda)\lambda^{2(n+1)}] \to [1 + \gamma \lambda^{n+2}] \quad {\rm for} \quad \lambda \to 0\\ \\
F'(\lambda) F(\lambda) + [\lambda F'(\lambda) - (n+1)F(\lambda)]G(\lambda)\lambda^{2n+1} = 0 \, .
\end{array}
\right.
\label{sysFG}
\ee
We have to see if the two conditions are compatible, and what this implies for $f$ and $g$.
To check this compatibility we can use power series representations of the functions $F(\lambda)$,
$G(\lambda)$. We allow $G(\lambda)$ to develop poles. Since the factor $\lambda^{2(n+1)}$ multiplies
$G(\lambda)$ in the boundary condition, we could allow poles until $\lambda^{-2(n+1)}$ and still the
combination $[F + G\lambda^{2(n+1)}]$ would remain analytical. However, we'll show that the result can
be obtained by allowing poles just until $\lambda^{-n}$ only. So we write
\be \left\{ \begin{array}{ll}
F(\lambda) = \sum_{k=0}^{\infty} a_k \lambda^k\\ \\
G(\lambda) = \sum_{k=-n}^{\infty} b_k \lambda^k \, .
\end{array}
\right.
\ee
and we look for what the two conditions imply on the coefficients $a_k$, $b_k$. We have
\be
F(\lambda) + G(\lambda)\lambda^{2(n+1)} = \sum_{k=0}^{\infty}(a_k + b_{k-2(n+1)})\lambda^k
\label{FGL}
\ee
where $b_{k-2(n+1)}=0$ for $k=0,1,2,...,n+1$ and $b_{-n}\neq0$, $b_{-n+1}\neq0$, etc.\\
At small $\lambda$ we should have the matching, for $\lambda \to 0$,
\be
\sum_{k=0}^{\infty}(a_k + b_{k-2(n+1)})\lambda^k \longrightarrow 1 + \gamma \lambda^{n+2}\, .
\label{condlim}
\ee
This means
\be
k&=&0; \quad \quad [a_0 + b_{-2(n+1)}] = 1 \quad \Rightarrow \quad a_0=1 \nonumber \\
k&=&1; \quad \quad [a_1 + b_{1-2(n+1)}] = 0 \quad \Rightarrow \quad a_1=0 \nonumber \\
k&=&2; \quad \quad [a_2 + b_{2-2(n+1)}] = 0 \quad \Rightarrow \quad a_2=0 \nonumber \\
&...& \quad \quad \quad \quad... \quad \quad \quad \quad ...  \nonumber \\
k&=&n+2; \quad \quad [a_{n+2} + b_{-n}] = \gamma \quad \Rightarrow \quad (*)\nonumber \\
k&=&n+3; \quad \quad [a_{n+3} + b_{-n+1}] = {\rm any \,\,\, quantity} \nonumber \\
&...& \quad \quad \quad \quad... \quad \quad \quad \quad ...
\label{relab}
\ee
(*) here at least $b_{-n}$ is $\neq 0$, therefore at least $b_{-n}$ can be chosen $=\gamma$.

Now let's see if the conditions on $a_k$, $b_k$ just found above are compatible with those required
by the differential equation (\ref{sysFG}). Since
\be
F'(\lambda)&=& \sum_{k=0}^{\infty}(k+1)a_{k+1}\lambda^k \nonumber \\
G(\lambda)&=& \sum_{k=-n}^{\infty}b_k\lambda^k = \sum_{k=0}^{\infty}b_{k-n}\lambda^{k-n}
\ee
with $b_{-n}\neq0$, $b_{-n+1}\neq0$,..., $b_0\neq0$, we have
\be
F(\lambda)F'(\lambda)&=&\sum_{k=0}^{\infty}C_k\lambda^k \nonumber \\
G(\lambda)F'(\lambda)&=&\sum_{k=0}^{\infty}D_k\lambda^{k-n} \nonumber \\
F(\lambda)G(\lambda)&=&\sum_{k=0}^{\infty}E_k\lambda^{k-n} \nonumber
\ee
where
\be
C_k&=&\sum_{q=0}^{k}(q+1)a_{k-q}a_{q+1} \nonumber \\
D_k&=&\sum_{q=0}^{k}(q+1)b_{k-n-q}a_{q+1} \nonumber \\
E_k&=&\sum_{q=0}^{k}a_{k-q}b_{q-n}
\ee
and Eq.(\ref{sysFG}) becomes
\be
\sum_{k=0}^{\infty}\left[C_k \lambda^k + D_k \lambda^{k+n+2} - (n+1)E_k\lambda^{k+n+1}\right] = 0
\label{CDE}
\ee
Reshuffling indexes a bit in Eq.(\ref{CDE}) we get
\be
\sum_{k=0}^{\infty}\left[C_k + D_{k-n-2} - (n+1)E_{k-n-1}\right]\lambda^k = 0
\label{CDE2}
\ee
where
\begin{eqnarray}
  D_{-n-2}&=&0 \quad \quad \quad E_{-n-1}=0 \nonumber \\
  D_{1-n-2}&=&0 \quad \quad \quad E_{1-n-1}=0 \nonumber \\
  ... &\quad& ... \quad \quad \quad ... \quad\quad ...\nonumber \\
  D_{-1}&=&0 \quad \quad \quad E_{-1}=0 \nonumber
\end{eqnarray}
and it is easy to see that these relations are direct consequences of the definitions for $b_k$
in Eq.(\ref{FGL}) and of relations (\ref{relab}). Equation (\ref{CDE2}) can be satisfied only if all
the coefficients of $\lambda^k$ are identically zero. We can now check explicitly that this requirement is
in full agreement with conditions (\ref{relab}). In fact
\begin{eqnarray}
&&k=0; \nonumber \\
&&C_0 + D_{-n-2}-(n+1)E_{-n-1}=a_0a_1=0 \,\,\Rightarrow\,\, a_1=0 \nonumber
\end{eqnarray}
(since $a_0=1$) and this agrees with (\ref{relab}). And then
\be
&&k=1; \nonumber \\
&&C_1 + D_{1-n-2}-(n+1)E_{1-n-1}=a_1a_1+2a_0a_2=0 \nonumber \\
&& \Rightarrow\,\, a_2=0 \nonumber
\ee
(since $a_0=1$) and this agrees with (\ref{relab}). Again, for $k=2$ we have
\be
C_2 + D_{2-n-2}-(n+1)E_{2-n-1}=0 \,\,\Rightarrow\,\, a_3=0 \nonumber
\ee
and so on for $k=3,4,...$.\\
For $k=n$ we find $a_{n+1}=0$ in agreement with (\ref{relab}).\\
For $k=n+1$ we have
\be
&&C_{n+1}+D_{-1}-(n+1)E_0 \nonumber \\
&&=\sum_{q=0}^{n+1}(q+1)a_{n+1-q}a_{q+1} - (n+1)\sum_{q=0}^{0}a_{0-q}b_{q-n}\nonumber \\
&&=(n+2)a_0a_{n+2} - (n+1)a_0b_{-n} = 0
\ee
Since $a_0=1$, then
\be
(n+2)a_{n+2} - (n+1)b_{-n} = 0
\ee
and this equation $is$ compatible with the "$k=n+2$" condition of (\ref{relab}).
In fact, we have two equations in two unknowns
\be \left\{ \begin{array}{ll}
(n+2)a_{n+2} - (n+1)b_{-n} = 0\\ \\
a_{n+2} + b_{-n} = \gamma \, .
\end{array}
\right.
\ee
which allow us to compute $a_{n+2}$ (the first non zero coefficient for $F(\lambda)$, after $a_0=1$)
and $b_{-n}$ (pole of order $n$ of $G(\lambda)$).\\
For the next case, $k=n+2$, we don't have evidently any problem, since Eq.(\ref{relab}) simply gives
$(a_{n+3} + b_{-n+1})=$\emph{any quantity}. Therefore any relation between $a_{n+3}$, $b_{-n+1}$
required by the differential equation in (\ref{sysFG}) is acceptable.
Note moreover that if we allowed poles for $G(\lambda)$ with a degree less than $n$,
we would find contradiction between the conditions (\ref{condlim}), (\ref{relab}) and the
differential equation in (\ref{sysFG}).

Thus, we conclude that the two conditions (\ref{sysFG}) are compatible (if we allow $G(\lambda)$ to develop
poles). So the MBH GUP is translational and rotational invariant. Q.E.D.
%
%
%
\section*{Appendix 2}
%
%
%
In this Appendix we show that the analytic form of the relation $E(T)$ does not affect the minimum masses
computable from (\ref{ME}), provided that $E$ is a monotonically increasing function of $T$. In fact,
the mass-energy formulae (\ref{ME}) can be written, in general, as
\be
A(m)= \frac{1}{E(T)} + \lambda E(T)^\eta =: f(T)
\label{AM}
\ee
where $A(m)$ is supposed to be a monotonically increasing function of the mass $m$ and $\lambda>0$, $\eta>0$.
Then, to get the minimum mass predicted by (\ref{AM}), it is sufficient to compute
\be
f'(T)=\left[\lambda \eta E(T)^{\eta -1} - \frac{1}{E(T)^2}\right]E'(T)=0\,. \nonumber
\ee
Discarding the trivial solution $E'(T)=0$, we get
\be
E(T_c)=\left(\frac{1}{\eta \lambda}\right)^{\frac{1}{\eta + 1}} \quad \Rightarrow \quad
T_c=E^{-1}\left[\left(\frac{1}{\eta \lambda}\right)^{\frac{1}{\eta + 1}}\right]\,. \nonumber
\ee
Then
\be
A(m_{MIN})=\frac{1}{E(T_c)} + \lambda E(T_c)^\eta =
\left(\frac{\eta+1}{\eta} \right)(\eta \lambda)^{\frac{1}{\eta+1}} \nonumber
\ee
and
\be
m_{MIN}=A^{-1}\left[  \left(\frac{\eta+1}{\eta} \right)(\eta \lambda)^{\frac{1}{\eta+1}}  \right]\,. \nonumber
\ee
As we see, the minimum mass, a very relevant prediction of relations like (\ref{ME}), does not depend
in any way from the explicit form of $E(T)$. Q.E.D.

%
%
%


\end{document}